\documentclass[11pt]{article}
\usepackage{jheppub}
\usepackage{amsmath,amssymb,amsfonts,graphicx,slashed,color}
\usepackage{subcaption}
\usepackage{tikz}
\usepackage{tikz-3dplot}
\usepackage{pgfplots}
\usepgfplotslibrary{colormaps}
\usetikzlibrary{decorations.markings}
\usetikzlibrary{decorations.pathmorphing}
\usepgfplotslibrary{fillbetween}
\usetikzlibrary{patterns}
\usepackage[utf8]{inputenc}
\usetikzlibrary{decorations.markings}
\usetikzlibrary{decorations.pathmorphing}
\newlength{\lx}
\newlength{\ly}
\newcommand{\be}{\begin{equation}}
\newcommand{\ee}{\end{equation}}
\newcommand{\bea}{\begin{eqnarray}}
\newcommand{\eea}{\end{eqnarray}}
\newcommand{\beq}{\begin{equation}}
\newcommand{\eeq}{\end{equation}}
\newcommand{\beqn}{\begin{eqnarray}}
\newcommand{\eeqn}{\end{eqnarray}}

\newcommand{\cO}{\mathcal{O}}

\usepackage{upgreek}

\title{Semi-classical BMS$_3$ blocks and flat holography}

\author{Eliot Hijano}

\affiliation{Department of Physics and Astronomy, University of British Columbia,\\
6224 Agricultural Road, Vancouver, B.C.\ V6T 1Z1, Canada.}

\emailAdd{ehijano@phas.ubc.ca}

\abstract{
We present the construction of  BMS$_3$ blocks in a two-dimensional field theory  and compare the results with holographic computations involving probe particles propagating in flat space cosmologies. On the field theory side, we generalize the monodromy method used in the context of AdS/CFT to theories with BMS symmetry. On the bulk side, we consider geodesic Feynman diagrams, recently introduced in \cite{Hijano:2017eii}, evaluated in locally flat geometries generated by backreaction of heavy BMS primary operators. We comment on the implications of these results for the eigenstate thermalization hypothesis in flat holography.
}

\keywords{}

\begin{document}
\tikzset{->-/.style={decoration={
  markings,
  mark=at position #1 with {\arrow{>}}},postaction={decorate}}}

\def \L {10}
\def \H {1.5*\L}

\tikzset{
    mark position/.style args={#1(#2)}{
        postaction={
            decorate,
            decoration={
                markings,
                mark=at position #1 with \coordinate (#2);
            }
        }
    }
}

\tikzset{point/.style={insert path={ node[scale=2.5*sqrt(\pgflinewidth)]{.} }}}

\tikzset{->-/.style={decoration={
  markings,
  mark=at position #1 with {\arrow{>}}},postaction={decorate}}}

  \tikzset{-dot-/.style={decoration={
  markings,
  mark=at position #1 with {\fill[red] circle [radius=3pt,red];}},postaction={decorate}}} 

 \tikzset{-dot2-/.style={decoration={
  markings,
  mark=at position #1 with {\fill[blue] circle [radius=3pt,blue];}},postaction={decorate}}} 

 \tikzset{-dotg-/.style={decoration={
  markings,
  mark=at position #1 with {\fill[green] circle [radius=3pt,green];}},postaction={decorate}}} 
  
   \tikzset{-dotp-/.style={decoration={
  markings,
  mark=at position #1 with {\fill[violet] circle [radius=3pt,violet];}},postaction={decorate}}} 
  
     \tikzset{-dotb-/.style={decoration={
  markings,
  mark=at position #1 with {\fill[black] circle [radius=3pt,black];}},postaction={decorate}}} 
  
\tikzset{snake it/.style={decorate, decoration=snake}}
    
    \definecolor{darkgreen}{RGB}{0,180,0}


\maketitle

\parskip=10pt

\section{Introduction}

The application of the holographic principle \cite{tHooft:1999rgb,Susskind:1994vu} to asymptotically flat space-times  implies that theories of quantum gravity in flat blackgrounds are dual to field theories with BMS symmetry (BMSFT) \cite{Bagchi:2010eg,Bagchi:2012cy,Barnich:2010eb}.  BMS symmetry is the asymptotic symmetry group of asymptotically flat geometries at null infinity \cite{Bondi:1962px,Sachs:1962zza}. In this paper we explore the three dimensional case, which provides a testing ground for ideas concerning classical and quantum gravity in flat space-times. Our objective is two-fold; first, to further test the extrapolate dictionary proposed in \cite{Hijano:2017eii}, where it was conjectured that position-space Feynman diagrams integrated over null geodesics in Minkowski space compute BMSFT correlators of local BMS primaries. Second, to make a precise statement concerning the eigenstate thermalization hypothesis in the context of flat space holography. 

It has been hypothesized that the high energy microstates of a quantum system behave as a thermal background \cite{Srednicki:1995pt,PhysRevA.43.2046}. The set of ideas explaining when and why this is true is known as the eigenstate thermalization hypothesis (ETH).  Much of the recent research concerning ETH is in the context of conformal field theories and AdS/CFT \cite{Lashkari:2016vgj,Fitzpatrick:2016mjq,Fitzpatrick:2015zha,Faulkner:2017hll}.   A more concrete version of ETH is that expectation values of few-body observables in a highly energetic microstate match those of the micro-canonical ensemble up to corrections that are exponentially suppressed in entropy.  Throughout this work, we aim to test this hypothesis by computing observables in field theories expected to be dual to a theory of quantum gravity in three dimensional flat space.

Precise statements concerning ETH in the context of AdS/CFT were made in the illuminating work \cite{Fitzpatrick:2015zha}. In that work, Virasoro conformal blocks involving two heavy operators (high energy) where computed in the large central charge limit. The results were also matched to bulk processes where light particles interact with a thermal semi-classical background. This immediately implies that some form of ETH holds universally in two-dimensional CFT's. 

In this paper we explore the computation of BMS$_3$ blocks  both in the field theory and in the bulk. BMSFT correlators can be expanded as a sum over exchanges of the irreducible representations of the BMS algebra. The BMS blocks are expected to be an essential ingredient in the recently introduced BMS bootstrap program \cite{Bagchi:2017cpu,Bagchi:2016geg}, whose objective is to constraint the space of physically sensible field theories with BMS symmetry. In the field theory, we generalize the monodromy method used in conformal field theories to theories with BMS symmetry. We perform the computation in two different ways; on one hand, we set up the monodromy problem by finding the finite coordinate transformations that effectively place the field theory on a non-trivial background metric that replaces the operators in the correlator. On the other hand, we set up the monodromy problem by finding the null vectors of the BMS$_3$ algebra. It will turn out that the usual null vectors built out of irreducible representations of the symmetry algebra do not provide with a non-trivial monodromy problem. This will lead to the introduction of BMS multiplets; reducible but indecomposable representations of the BMS algebra, much like the ones appearing in the context of logarithmic CFT's \cite{Gurarie:1993xq}. 

In the bulk, we will match the field theory results to computations involving geodesic networks evaluated in non-trivial asymptotically flat geometries. In previous work \cite{Hijano:2017eii}, we computed Poincar\'e blocks (global BMS blocks) holographically. The computation consisted on extremizing the length of a geodesic network living in Minkowski space-time. Inspired by this construction, we proposed an extrapolate dictionary between correlators in a BMSFT and Feynman diagrams in the bulk. The computations presented here apply this dictionary to more general asymptotically Minkowski backgrounds.

The paper is organized as follows. In section \ref{sec:summary} we provide with a quick summary of our results. Section \ref{sec:BMSFTtech} develops some of the technology required to perform the field theory computations, while sections \ref{sec:BMSFT} and \ref{sec:null} solve the monodromy problem in two different ways. This results in an explicit closed form expression for BMS$_3$ blocks. Section \ref{sec:Holo} focuses on the holographic computations. We conclude with a discussion in section \ref{sec:discussion}.

\subsection{Summary of results}\label{sec:summary}
We study correlation functions of four local BMS$_3$ primary operators. These operators are labeled by two quantum numbers $\xi$ and $\Delta$, which are the eigenvalues associated to the elements of the center of the symmetry algebra.  Two of the operators are heavy (H), in the sense that they scale freely with the central charge of the theory $c_M$. The other two operators are light (L), which means that the quantum numbers obey $1\ll \xi_L,\Delta_L \ll c_M$, which we will refer to as the ``probe limit". The four-point function can be written as a sum over BMS invariant functions by inserting the identity operator as a sum over a complete set of states, which are organized in irreducible representations of the symmetry algebra
\be
{\cal F}=\sum_{\alpha} \langle H(x_1) H(x_2) \vert \alpha \rangle\langle \alpha \vert L(x_3) L(x_4) \rangle=\sum_{\alpha} {\cal F}_{\alpha}\, .
\ee
The operators are located at points $x_i$ in the null plane, which is parametrized by the plane coordinates $u_i$ and $v_i$. We consider blocks where the exchanged representation $\alpha$ is also light. The main technical result of this paper is a closed form expression for the block, which we anticipate here
\be\label{eq:SummaryResult1}
\begin{split}
{\cal F}_{\alpha} &=   \left( {{   U^{\beta-1}    }\over{  (1-U^{\beta})^2    }}   \right)^{\Delta_L} e^{V\left(  {{\beta U^{{{\beta}\over 2}}}\over{U(U^{\beta}-1)}}\xi_{\alpha} - {{U^{\beta}(\beta+1)+\beta-1}\over{U(U^{\beta}-1)}}\xi_L \right)
}\, \\
&\times  \left( {{1-U^{{\beta}\over{2}}}\over{1+U^{{\beta}\over{2}}}}  \right)^{\Delta_{\alpha}}  e^{\Delta_H \log U \left(     {{2U^{{\beta}\over 2}}\over{\beta (U^{\beta}-1)}}    \xi_{\alpha}            +  {{2(U^{\beta}+1)}\over{\beta(1-U^{\beta})}} \xi_L                    \right)
}\, .
\end{split}
\ee
Here, we have introduced the parameter $\beta=\sqrt{1-24\xi_H/c_M}$, as well as the BMS cross-ratios
\be
U={{u_{23}u_{14}}\over{u_{13}u_{24}}}\, , \quad \text{and} \quad {V\over U}={{v_{13}}\over{u_{13}}}+{{v_{34}}\over{u_{34}}}-{{v_{12}}\over{u_{12}}}-{{v_{34}}\over{u_{34}}}\, .
\ee
The result \ref{eq:SummaryResult1} will be obtained in the field theory by studying  null vectors of the BMS algebra, as well as the geometric monodromy method. We will also show that the result is consistent with the non-relativistic limit of a Virasoro block.

The holographic computation follows the prescription introduced in \cite{Hijano:2017eii}, with a new ingredient; the heavy operators in the correlator correspond to a different asymptotically Minkowski background geometry. The computation consists on integrating a Feynman diagram over the null lines falling from the boundary at the locations of the light operators. In the proble  limit, such computation can be well approximated by a stationary phase approximation, which leads to the extremization of a geodesic network. We will consider background geometries known as flat space cosmological solutions (FSC). These solutions are quotients of Minkowski space and they are labeled by their mass $M$ and angular momentum $J$. The computations presented throughout this note match the field theory results anticipated in \ref{eq:SummaryResult1} upon the identification 
\be\label{eq:MJheavy}
M={{24\xi_H}\over{c_M}}-1\, , \quad  \text{and}\quad {J\over{2\sqrt{M}}}={{6\Delta_H}\over{c_M}}\, .
\ee
We conclude that the micro-state associated to a heavy operator leads to the same correlators found in a thermal background, which can be understood as the statement of the eigenstate thermalization hypothesis.

\section{BMS$_3$ field theory}\label{sec:BMSFTtech}
The technical objective of this work is to advance the recently introduced BMS bootstrap programme by explicitly computing four-point function blocks in a two-dimensional theory with BMS$_3$ symmetry. The field theory results  will then be analyzed holographically, aiming to expand our understanding of the holographic dictionary relating field theories with BMS symmetry and gravitational theories in asymptotically flat space-times. The main technical tool we will present here is a flat version of the monodromy method. The monodromy method has been used in CFT$_2$ to compute Virasoro blocks (see for example \cite{Harlow:2011ny,Fitzpatrick:2015zha,Anous:2016kss}) and generalizations such as higher spin blocks \cite{deBoer:2014sna}.  In the following subsections we set up some basic formalism that will be used throughout this work. Namely, we will derive how primaries of the algebra and currents transform under finite BMS$_3$ transformations, as well as the Ward identities concerning expectation values of the conserved currents of the theory. 

\subsection{The BMS$_3$ algebra}
The BMS$_3$ algebra is a central extension of the algebra obeyed by surface charges associated with the symmetries of asymptotically flat three dimensional spacetimes at null infinity. If we choose global Minkowski coordinates, the line element of the bulk geometry reads
\be
ds^2 = -d\tau^2 -2d\tau dr +r^2 d\phi^2\, ,
\ee
where $\tau=t-r$ is the retarded time coordinate. In this work we are concerned with the asymptotic symmetries at null infinity, which is located at $r\rightarrow \infty$, with $\tau$ and $\phi$ fixed. The asymptotic killing vectors at the null boundary can be explicitly written as
\be\label{eq:CylRep}
l_a = e^{i a \phi} \left(   \partial_{\phi}  +i a \tau  \partial_{\tau}   \right)\, , \quad \text{and} \quad m_a=e^{i a \phi} \partial_{\tau}\, .
\ee
The commutators of these vectors obey the following algebra 
\be\label{eq:algebra}
\begin{split}
[l_a,l_b]&=(a-b)l_{a+b}\, , \\
[l_a,m_b]&=(a-b)m_{a+b}\, , \\
[m_a,m_b]&=0\, .
\end{split}
\ee
The formulas \ref{eq:CylRep} are known as the cylinder representation of the algebra \ref{eq:algebra}. The reason for this is the cylinder coordinates choosen to parametrise global Minkowski space-time. Throughout this work, we will also make use of the ``plane representation'' of the algebra given by
\be\label{eq:geometric}
l_a = -u^{a+1}\partial_u - (a+1)u^a v \partial_v\, , \quad \text{and} \quad m_a=u^{a+1}\partial_v\, .
\ee
The plane and cylinder representations are related by the following map
\be\label{eq:map}
u=e^{i \phi}\, , \quad \text{and} \quad v=i \tau e^{i \phi}\, .
\ee
In this note, we will find it convenient to perform the computations in the BMSFT using the plane representation. However, holographic results are more naturally understood in the cylinder one, as it is directly connected to global Minkowski coordinates. The map \ref{eq:map} will be used to compare results in both sides of the duality.

The algebra spelled out in \ref{eq:algebra} admits a central extension
\be\label{eq:algebraC}
\begin{split}
[L_a,L_b]&=(a-b)L_{a+b}+{{c_L}\over{12}}a(a^2-1)\delta_{a+b,0}\, , \\
[L_a,M_b]&=(a-b)M_{a+b}+{{c_M}\over{12}}a(a^2-1)\delta_{a+b,0}\, , \\
[M_a,M_b]&=0\, .
\end{split}
\ee
The central charges $c_L$ and $c_M$ depend on the specific theory of gravity in the bulk. For the case of Einstein gravity,
in order for the phase space of three-dimensional asymptotically flat gravity to match the
space of coadjoint representations of the BMS$_3$ group, it is required that \cite{Barnich:2006av,Barnich:2015uva,Oblak:2016eij}
\be
c_L=0\, \quad \text{and }\quad c_M={3\over G}\, .
\ee
We turn now to the study of how finite BMS$_3$ transformations act on the primaries and currents of the field theory.

\subsection{Finite BMS$_3$ transformations} 
The geometric approach to the monodromy method will  require us to understand how finite BMS transformations act on BMSFT primaries and currents. For example we would like to obtain an explicit expression for
\be\label{eq:objective}
\phi'(x')=e^{\sum_a\omega_a T^a}\phi(x) \, ,
\ee
where $\omega_a$ are arbitrary parameters defining the finite transformation and $T_a$ are the generators of the symmetry algebra. The strategy is the following; we will first define the subalgebra of generators that leave the origin invariant, and define a matrix representation for the generators of this subalgebra at the origin. The action of these matrices on primaries at the origin is given by the fact that primaries are highest weights in irreducible representations of the subalgebra. We will then apply the Beker-Campbell-Hausdorff formula (BCH) to derive the action of the generators of the symmetry algebra away from the origin. Exponentiating the generators will then lead to an explicit expression for \ref{eq:objective}.

 The starting point is to consider the subgroup of the BMS$_3$ algebra that leaves the origin   invariant. Looking at the geometric realization \ref{eq:geometric}, it is clear that the point $(u,v)=0$ is invariant under the action of $l_a$ and $m_a$ with $a\ge 0$. These generators necessarily form a closed subalgebra, and their action on the primaries can be re-named by introducing a matrix representation that defines the action of BMS generators on the field $\phi(0)$ at the origin.
\be
l_{a\ge 0}\phi(0) \equiv {\mathfrak{l}}_a \phi(0)\, , \quad \text{and} \quad m_{a\ge 0}\phi(0) \equiv {\mathfrak{m}}_a \phi(0)\, .
\ee
The field $\phi(0)$ corresponds to an irreducible highest weight representation of the subalgebra spanned by the generators ${\mathfrak{m}}_a$ and ${\mathfrak{l}}_a$. The action of the BMS generators on the field at a point away from the origin can be derived through the BCH formula. Namely,
\be\label{eq:Hausdorff}
e^{-A} B e^{A} = B+[B,A]+{1\over {2!}}[[B,A],A]+\dots
\ee
with $A=i x^{\rho}P_{\rho}$ and $B$ are the generators ${\mathfrak{l}}_a$ or ${\mathfrak{m}}_a$. A subtlety here is that the vector $x^{\rho}$ is  a vector of c-numbers and not an operator like $P_{\rho}$ or ${\mathfrak{l}}_a$ and ${\mathfrak{m}}_a$. We thus write
\be
i x^{\rho}P_{\rho} = -U l_{-1} +V m_{-1}\, ,
\ee
where capital letters commute with derivative operators. We can now compute the action of the generators ${\mathfrak{l}}_a$ and ${\mathfrak{m}}_a$ away from the origin by applying the BCH formula and making use of the commutators that define the BMS$_3$ algebra. Here we will only present the computation involving the generators ${\mathfrak{m}}_a$. Using the commutation relations \ref{eq:algebra} we obtain the following expression for the $p$-nested commutator of $m_n$ with $ix^{\rho}P_{\rho} $
\be
[[[m_n, ix^{\rho}P_{\rho} ],\dots ],ix^{\rho}P_{\rho}]=\left( -U \right)^p {{(n+1)!}\over{(n+1-p)!}}m_{n-p}\, .
\ee
The BCH formula then implies
\be\label{eq:theta}
{\mathfrak{m}}_n\phi (U,V)  = \sum_{p=0}^{n+1}\left( -U\right)^p {{(n+1)!}\over {p! (n+1-p)!}} m_{n-p} \phi (U,V) \, .
\ee
We are concerned with the action of these generators on primary fields of the BMSFT. Such fields have an eigenvalue $\xi$ with ${\mathfrak{m}}_0$, and are killed by the positive modes ${\mathfrak{m}}_{>0}$. The relation \ref{eq:theta} can be inverted to obtain
\be\label{eq:thetainv}
 m_n \phi (U,V) = \xi U^n (n+1) \phi(U,V) +U^{n+1}m_{-1}\phi(U,V)   \, .
\ee
This result has been obtained previously in \cite{Bagchi:2009ca} in the context of Galilean conformal algebras. We are now ready to write the following expression for a finite BMS$_3$ transformation involving exclusively the generators $m_n$
\be\label{eq:PrimaryTransformation}
\begin{split}
\phi'(x')&=e^{\sum_a\omega_a \left(  \xi u^a (a+1) +u^{a+1}\partial_v  \right)}\phi(x) \, \\
&=e^{\xi G'(u)}\phi (u,v+G(u))\, ,
\end{split}
\ee
where we have introduced the function
\be
G(u)=\sum_a \omega_a u^a\, .
\ee
The same logic can be applied to the generators ${\mathfrak{l}}_a$, leading to the full expression
\be\label{eq:primTransf}
\phi'(x')=\left(P'\right)^{\Delta} e^{\xi \left(  {{G'}\over{P'}}+v{{P''}\over{P'}}   \right)}\phi (P,vP'+G)\, ,
\ee
where $P(u)$ and $G(u)$ are arbitrary functions of $u$ that depend on the gauge parameters $\omega_a$. It is clear then that a finite BMS transformation acts as  \ref{eq:primTransf} on primary fields, and on the coordinates as
\be\label{eq:CoordTransf}
u\rightarrow P(u)\, \quad \text{and}\quad  v\rightarrow v P'(u)+Q(u)\, .
\ee
Having dealt with how primary fields transform under finite BMS$_3$ transformations, we turn our attention to the conserved stress tensor.

\subsection{BMS$_3$ currents}\label{sec:currents}
Throughout this work it will be crucial to understand how the conserved currents of the theory transform under finite BMS$_3$ transformations. The strategy to writing the transformation laws will be very similar to the one used to study primary fields, with the complication that the central terms of the BMS algebra now play a role. We start by defining our stress tensor operator as an expansion in generators of the BMS$_3$ algebra. The generators $M_n$ and $L_n$ can be understood as the Noether charges associated with the symmetries of asymptotically flat three-dimensional geometries. In other words, the commutators in \ref{eq:algebraC} correspond to Poisson brackets involving Noether charges of the bulk theory in the Hamiltonian formulation.  The symmetries of the theory correspond to diffeomorphisms, which act on the coordinates as in equation \ref{eq:CoordTransf}. The Noether current is then related to the stress-energy tensor, such that
\be
Q_a = \int_{\Sigma} d\Sigma^{\mu} T_{\mu\nu} \xi^{\nu}_a \, ,
\ee
where $\Sigma$ is any complete space-like slice, and $\xi_a$ is the diffeomorphism implemented by the generator associated to $Q_a$. We will choose $\Sigma$ to be a slice of constant $\tau$ in cylindrical coordinates, which corresponds to a contour around the origin in the plane. We thus have
\be
Q_a = \oint du \, T_{u \nu} \xi^{\nu}_a \, .
\ee
We are now ready to write an expression for the generators $M_n$ and $L_n$;
\be\label{eq:MT}
\begin{split}
M_n&= \oint du\, T_{u v} u^{n+1}\, , \\
L_n&=-\oint du \, \left(   T_{u  u} u^{n+1} +T_{u v} (n+1) u^n v     \right)\, .
\end{split}
\ee
Inverting these formulas yields
\be\label{eq:TM}
\begin{split}
{\cal M}\equiv &\, T_{u v} = \sum_n M_n u^{-n-2}\, , \\
{\cal L}\equiv & \, T_{u u} = \sum_n \left[ L_n+(n+2){v \over u} M_n   \right] u^{-n-2}\, .
\end{split}
\ee
Where we have introduced the operators ${\cal M}$ and ${\cal L}$ to refer to the different components of the stress-energy tensor. In order to see what kind of operators we are working with, it is useful to write them at the origin by making use of the BCH formula \ref{eq:Hausdorff}.  For the operator $\cal M$, we have
\be
{\cal M}(0)={\mathfrak{m}}_{-2} \mathbf{1}(0)\, .
\ee
This object is not a primary in an irreducible representation of the BMS algebra, in the sense that it is not killed by positive modes of the algebra. More concretely, we have
\be
{\mathfrak{l}}_{n>0}{\cal M}(0) = {{c_M}\over{2}}\delta_{n,2}\mathbf{1}(0)\, , \quad \text{and} \quad {\mathfrak{m}}_{n>0}{\cal M}(0) = 0\, ,
\ee
where we are working with $c_L=0$. However, it is an eigenvector of ${\mathfrak{m}}_0$ and ${\mathfrak{l}}_0$
\be\label{eq:CenterM}
{\mathfrak{l}}_{0}{\cal M}(0) = 2 {\cal M}(0) \, , \quad \text{and} \quad {\mathfrak{m}}_{0}{\cal M}(0) = 0\, .
\ee
This implies that $\cal M$ is a quasi primary operator, with $\xi_{\cal M}=0$ and $\Delta_{\cal M}=2$. The other component of the stress tensor can be defined at the origin by
\be
{\cal L}(0)={\mathfrak{l}}_{-2} \mathbf{1}(0)\, .
\ee
Again, this is not a primary operator in an irreducible representation of the BMS algebra. The action of the positive modes is
\be
{\mathfrak{l}}_{n>0}{\cal L}(0) = 0\, , \quad \text{and} \quad {\mathfrak{m}}_{n>0}{\cal L}(0) = {{c_M}\over{2}}\delta_{n,2}\mathbf{1}(0)\, .
\ee
unlike ${\cal M}(0)$, ${\cal L}(0)$ is not an eigenvector of ${\mathfrak{l}}_0$ and ${\mathfrak{m}}_0$;
\be\label{eq:CenterL}
{\mathfrak{l}}_{0}{\cal L}(0) = 2{\cal L}(0)\, , \quad \text{and} \quad {\mathfrak{m}}_{0}{\cal L}(0) = 2{\cal M}(0) \, .
\ee
Formulas \ref{eq:CenterL} and \ref{eq:CenterM} suggest that the currents ${\cal M}$ and ${\cal L}$ form a ``BMS$_3$ multiplet''. We define multiplets as sets of operators whose action by the center of the algebra admits a Jordan cell structure. In other words, the action of $L_0$ and $M_0$ on the set of operators defining the multiplet is not diagonalizable. These multiplet structures will re-appear when discussing the null vectors of the BMS$_3$ algebra in section \ref{sec:null}.

This concludes the analysis of how the elements of the subalgebra spanned by $\mathfrak{m}_a$ and $\mathfrak{l}_a$  act on the operators $\cal M$ and $\cal L$ at the origin. We are now ready to apply the BCH formula and obtain expressions for the action of the generators of the symmetry algebra on the components of the stress tensor away from the origin.  For the component $\cal M$, we obtain
\be\label{eq:Maction}
\begin{split}
 m_n {\cal M}(U,V) &= U^{n+1}m_{-1}{\cal M}(U,V)   \, ,\\
l_n {\cal M}(U,V) &= {{c_M}\over{12}}(n^3-n)U^{n-2}\mathbf{1}(U,V)+U^n\left[2(n+1)+Ul_{-1}-(n+1)Vm_{-1}\right]{\cal M}(U,V)\, .
\end{split}
\ee
The first expression in \ref{eq:Maction} does not differ from the action of $m_n$ on a primary field, so the stress tensor ${\cal M}(U,V)$ transforms as a primary under gauge transformations generated by $m_n$. The second expression differs from the primary one by the term proportional to $c_M$. An infinitesimal transformation would read
\be\label{eq:Minf}{\cal M}'(x')={\cal M}(u,v)-{{c_M}\over{12}}\partial_u^3 \epsilon(u) -2\partial_u \epsilon(u) {\cal M}(u,v)-\epsilon(u) \partial_u {\cal M}(u,v)-v\partial_u\epsilon(u)\partial_v {\cal M}(u,v)\, .
\ee
Here $\epsilon(u)$ is a function related to the $P(u)$ we used for the primary field in the following way
\be
P(u)=u+\epsilon(u)
\ee
We are working linearly in $\epsilon(u)$, but the full answer can be obtained by working to higher orders, leading to
\be\label{eq:Minf}
{\cal M}'(x')=\left(  P' \right)^{2} {\cal M}(P,vP'+G) +{{c_M}\over{12}}S(P,u)\, .
\ee
Here we have introduced the Schwarzian of the coordinate transformation $P(u)$. The result for the component ${\cal L}$ requires a more cumbersome calculation, but the final result reads
\be
\begin{split}
{\cal L}'(x')=&\left( P' \right)^2 {\cal L}(P,vP'+G) +2P'(G'+vP''){\cal M}(P,vP+G)\\
&-{{c_M}\over{12}}{{G'\left(P'P''-3\left(P''\right)^3\right)+P'\left(3P''G''-P'G'''\right)}\over{\left(P'\right)^3}}\\
&+{{c_M}\over{12}}v{{3\left(P''\right)^2-4P'P''P'''+\left(P'\right)^2P'''}\over{\left(P'\right)^3}}\, .
\end{split}
\ee
It is possible to perform a check of this results through the limiting method. In appendix \ref{app:limit} we obtain the same formulas by taking a non-relativistic limit of the transformation laws present in two-dimensional theories with conformal symmetry.

\subsection{Ward identities}
In these notes we will make use of the Ward identities obeyed by correlators of primary operators with the conserved currents. Ward identities exist in any quantum field theory and a BMSFT is no different. In this section we will take the algebraic approach to such identities, instead of following the conventional approach concerning the study of the path integral  to compute correlators. We would like to obtain identities for the following expressions
\be
\begin{split}
{\cal M}(x,x_i)&=\langle {\cal M}(u,v)\phi (u_1,v_1)\dots \phi(u_n,v_n)\rangle\, ,\\
 {\cal L}(x,x_i)&=\langle {\cal L}(u,v)\phi (u_1,v_1)\dots \phi(u_n,v_n)\rangle\, .
 \end{split}
\ee
We start with the definition of the stress tensor operators
\be
{\cal M}(0)={\mathfrak{m}}_{-2} \mathbf{1}(0)\, , \quad \text{and} \quad {\cal L}(0)={\mathfrak{l}}_{-2} \mathbf{1}(0)\, .
\ee
We are planning to insert these operators at points different from the origin. We can use the BCH formula to write these operators at different points in the null plane. The result reads
\be
{\cal M}(u,v)=\sum_k u^k{\mathfrak{m}}_{-2-k} \mathbf{1}(0)\, , \quad \text{and} \quad {\cal L}(u,v)=\sum_k \left( u^k {\mathfrak{l}}_{-2-k} \mathbf{1}(0)  -k v u^{k-1}{\mathfrak{m}}_{-2-k} \mathbf{1}(0)  \right)\, .
\ee
We now want to perform the Wick contractions in our correlator. We start with the correlator involving $\mathcal{M}$.
\be
{\cal M}(x,x_i)= \sum^n_{i=1}\sum_k u^k \langle \mathbf{1}(0) \phi (u_1,v_1)\dots  \mathfrak{m}_{-2-k}\phi(u_i,v_i)   \dots\phi(u_n,v_n)\rangle \, .
\ee
We now use equation \ref{eq:thetainv}, which expresses the action of $\mathfrak{m}_n$ on a primary field at an arbitrary point in the null plane. 
\be
{\cal M}(x,x_i)= \sum^n_{i=1}\sum_k u^k \langle \mathbf{1}(0) \phi (u_1,v_1)\dots \left(\xi_i u_i^{-2-k} (-1-k) +u_i^{-1-k}\partial_{v_i}\right)\phi(u_i,v_i)   \dots\phi(u_n,v_n)\rangle \, .
\ee
Performing the sum over $k$ we arrive at the expression
\be\label{eq:WardM}
{\cal M}(x,x_i)= \sum^n_{i=1} \left({{\xi_i}\over{(u-u_i)^2}}+{1\over{u-u_i}}\partial_{v_i}\right)\langle  \phi (u_1,v_1)\dots\phi(u_n,v_n)\rangle \, .
\ee
The same kind of logic leads to the following expression for the other current
\be\label{eq:WardL}
{\cal L}(x,x_i)= \sum^n_{i=1} \left({{\Delta_i}\over{(u-u_i)^2}}-{1\over{u-u_i}}\partial_{u_i}+2\xi_i{{(v-v_i)}\over{(u-u_i)^3}}+{{v-v_i}\over{(u-u_i)^2}}\partial_{v_i}\right)\langle  \phi (u_1,v_1)\dots\phi(u_n,v_n)\rangle \, .
\ee
In the next sections we will make use of \ref{eq:WardM} and \ref{eq:WardL} and refer to them as BMS$_3$ Ward identities. Note that these formulae have been obtained previously in the literature in \cite{Bagchi:2009pe} by taking a non-relativistic limit of CFT$_2$ Ward identities. 

\section{Geometric monodromy method}\label{sec:BMSFT}
The main objective of this section is to construct BMS$_3$ blocks involving primary operators  whose quantum numbers scale with the central charge $c_M$ in such a way that the dual asymptotically flat geometry be backreacted. In previous work, we studied blocks with quantum numbers $\xi$, $\Delta$ $\sim$ $\epsilon c_M$ with $\epsilon\ll 1$. We will refer to these operators as ``$L$", for ``light".  We now will allow two out of the four operators in the correlator to scale with $c_M$ freely. We will denote such operators ``$H$", which stands for ``heavy".  The correlator we will be studying is then
\be
{\cal F}\equiv \langle H(x_1) H(x_2) L(x_3) L(x_4) \rangle\, .
\ee
We can insert the identity operator in between pairs of operators in order to express the correlator as a sum over products of three point functions. The identity operator concerns a sum over the complete set of states of the theory, which organize in irreducible representations of the BMS$_3$ algebra. Summing over all the terms in a given representation gives rise to an object dubbed the BMS$_3$ block. This results in an expansion of the correlator into a basis of BMS$_3$ invariant functions
\be\label{eq:decomp}
{\cal F}=\sum_{\alpha} \langle H(x_1) H(x_2) \vert \alpha \rangle\langle \alpha \vert L(x_3) L(x_4) \rangle=\sum_{\alpha} {\cal F}_{\alpha}\, ,
\ee
where $\alpha$ labels the different irreducible representations of the BMS$_3$ algebra. Figure \ref{fig:diagrams} shows a diagrammatic version of formula \ref{eq:decomp}.  The objective of this section is to compute the object ${\cal F}_{\alpha}$ using the geometric monodromy method studied in previous works in the AdS/CFT literature \cite{Faulkner:2017hll}. 
\begin{figure}[t!]
\centering
\begin{tabular}{lll}
{\huge   $ \langle H(x_1) H(x_2) L(x_3) L(x_4) \rangle = $} {\Huge  $  \sum\limits_{\alpha}$} & \raisebox{-.45\height}{             
 \begin{tikzpicture}[scale=1]

\draw[very thick, black] (-0.5,0) to (-1.41,1.41) node[left=4,above=1]{{\color{black}$H_1$}};
\draw[very thick, black] (-0.5,0) to (-1.41,-1.41) node[left=4,below=1]{{\color{black}$H_2$}};

\draw[very thick, black] (0.5,0) to (1.41,1.41) node[right=4,above=1]{{\color{black}$L_4$}};
\draw[very thick, black] (0.5,0) to (1.41,-1.41) node[right=4,below=1]{{\color{black}$L_3$}};

\draw[very thick, black] (-0.5,0) to (0.5,0) node[pos=0.5,above]{{ {\color{black}$\alpha$}}};
\end{tikzpicture}
                } \\
\end{tabular}
\caption{BMS block expansion of a four-point function. The heavy operators are labeled as ``H", while the light ones are labeled by ``L". The choice of channel corresponds to the two heavy operators interchanging a BMS highest weight representation with the light operators. The exchanged representation is labeled by $\alpha$.}\label{fig:diagrams}
\end{figure}

The method consists on finding the finite BMS transformation that turns on the stress tensor expectation values associated with the insertions of the operators present in the four-point function. Such transformation depends on ``auxiliary parameters" related to coordinate derivatives of the correlator. The resulting BMS transformation carries gauge invariant data; it consists of the monodromy of the BMS transformation when cycling around a particular pair of operators. The choice of monodromy is related to the choice of exchanged representation $\alpha$. Demanding that the transformation has a particular monodromy fixes the auxiliary parameters, and in turn fixes the BMS block. We start by considering the monodromy of the three-point function.

\subsection{Three-point function monodromy}
We start by studying the relevant monodromy in the context of a three point function. The reason to do this is that we would like to replace a pair of operators in the four point function by a single family labeled by $\alpha$. This means that the monodromy of the gauge transformation when cycling around the pair of operators has to be equivalent to the monodromy of the three point function when circling around a single operator. We will thus be looking at the correlator
\be
\langle H(x_1) H(x_2) L_{\alpha} (x_{\alpha}) \rangle \, .
\ee
The index $\alpha$ labels the primary of the exchanged irreducible representation appearing in \ref{eq:decomp}. The expectation value of the stress tensors in the presence of the three operators can be obtained using the Ward identities. Focusing firs on ${\cal M}$ we obtain
\be
\begin{split}
{\cal M}(x_1,x_2,x_{\alpha};x)&={{\langle {\cal M}(x) H(x_1) H(x_2) L_{\alpha} (x_{\alpha}) \rangle }\over{\langle H(x_1) H(x_2) L_{\alpha} (x_{\alpha}) \rangle }}   \\
&=\sum^3_{i=1} {{ \left({{\xi_i}\over{(u-u_i)^2}}+{1\over{u-u_i}}\partial_{v_i}\right) \langle H(x_1) H(x_2) L_{\alpha} (x_{\alpha}) \rangle }\over{\langle H(x_1) H(x_2) L_{\alpha} (x_{\alpha}) \rangle }}
\end{split}
\ee
if we define the auxiliary parameters 
\be
c_i={6\over {c_M}}\partial_{v_i} \log \langle {\cal M}(x) H(x_1) H(x_2) L_{\alpha} (x_{\alpha}) \rangle\, ,
\ee
then we can write
\be
{\cal M}(x_1,x_2,x_{\alpha};x)=\sum^3_{i=1}  \left({{\xi_i}\over{(u-u_i)^2}}+{{c_M}\over 6}{{c_i}\over{u-u_i}}\right) \, .
\ee
The expression for the BMS$_3$ invariant three point function is known. See for example \cite{Bagchi:2017cpu}. Using the explicit expression one can compute the objects $c_i$. Using freedom to change coordinates, we place the operators at $u_1=0$, $u_2=\infty$, and leave $u_3$ as a free parameter. This results in
 \be
{\cal M}(u_3;u)={1\over{u^2}}\xi_H +{{u_3}\over{u(u-u_3)^2}}\xi_{\alpha} \, .
\ee
We are now looking for a coordinate transformation that takes the plane to a geometry with the same expectation value of the current. Looking at equation \ref{eq:Minf} implies the following differential equation that we need to solve
 \be\label{eq:Peq}
{{c_M}\over{12}}S(P,u)={1\over{u^2}}\xi_H +{{u_3}\over{u(u-u_3)^2}}\xi_{\alpha} \, .
\ee
We now define the following objects
 \be
\beta=\sqrt{1-24{{\xi_H}\over{c_M}}}\, , \quad \text{and} \quad \epsilon_{\alpha}=6{{\xi_{\alpha}}\over{c_M}}\, ,
\ee
and solve the problem to linear oder in the light operator dimension $\epsilon_{\alpha}$. Note that the differential equation can be re-casted as follows
 \be\label{eq:diffg}
g''(u) +{1\over 2}S(P,u) g(u)=g''(u) +{6\over {c_M}}{\cal M}(u_3;u) g(u) =0\, ,
\ee
where $P=g_1/g_2$, and $g_i$ are the two solutions of \ref{eq:diffg}. To zeroth order in $\epsilon_{\alpha}$, we obtain
 \be
g_1^{(0)}=u^{{{1+\beta}\over{2}}}\, , \quad \text{and} \quad g_2^{(0)}=u^{{{1-\beta}\over{2}}}\, .
\ee
In order to obtain the solutions to linear order in $\epsilon_{\alpha}$, we expand all of the functions involved in the differential equation;
\be\label{eq:AppAeq}
\begin{split}
g_i&=g_i^{(0)}+\epsilon_{\alpha} g_i^{(1)}\, , \\
{6\over {c_M}}{\cal M}(u_3;u)&={1\over{u^2}}{{1-\beta^2}\over 4} -{{u_3}\over{u(u-u_3)^2}}\epsilon_{\alpha}={6\over {c_M}}{\cal M}^{(0)}(u_3;u)+\epsilon_{\alpha}{6\over {c_M}}{\cal M}^{(1)}(u_3;u)  \, .
\end{split}
\ee
Expanding the differential equation itself, we obtain
\be\label{eq:diffO13}
g_i^{(1)\prime\prime}+{6\over {c_M}}{\cal M}^{(0)}(u_3;u)g_i^{(1)}=-{6\over {c_M}}{\cal M}^{(1)}(u_3;u)g_i^{(0)}\, .
\ee
This is now an inhomogeneous differential equation whose solution reads
\be
g_i^{(1)}(u)=f_{i,1}(u)g_1^{(0)}(u)+f_{i,2}(u)g_2^{(0)}(u)\, ,
\ee
where the functions $f_{i,j}$ obey the following equalities
\be\label{eq:fprimeM}
f^{\prime}_{i,j}(u)={{W_{i,j}(u)}\over{W(u)}}\, .
\ee
Here $W(u)$ is the Wronskian of the family of solutions, and $W_{i,j}(u)$ is the Wronskian with the $j^{\text{th}}$ row replaced by $\{0, {6\over {c_M}}{\cal M}^{(1)}(u_3;u)g_i^{(0)}\}$. In order to obtain the full solution we would have to integrate the expressions \ref{eq:fprimeM}. However, in order to compute the monodromy of $P(u)$ around the light operator, equation \ref{eq:fprimeM} is enough. 

So far we have been occupied figuring out what coordinate transformation takes the plane to a geometry whose expectation values of the currents matches the ones one would find in presence of two heavy and one light operators in the null plane. We now want to understand what happens to this coordinate transformation when we take the current around a loop containing the light operator. Doing this does not change the expectation value of the current, and so we still expect the function $P(u)$ to solve the differential equation \ref{eq:Peq}. However, each of the solutions $g_i$ can turn into a linear combination of $g_i$'s. This means that cycling around the light operator can result in the following transformation of $P$
\be
P(u)={{g_1(u)}\over{g_2(u)}} \xrightarrow{C_{\alpha}} {{A g_1(u) +B g_2(u)}\over{C g_1(u)+D g_2(u)}}={{A P(u)+B}\over{C P(u) +D}}\, .
\ee
The constants $A,B,C,$ and $D$ can be obtained by looking at the branch structure of the functions $f_{i,j}$. In more practical words, one can obtain these objects by computing the residue of the integral defining $f_{i,j}$ in equation \ref{eq:fprimeM} at $u=u_3$. However, not all of these constants are physically meaningful. Instead of simply circling around the light operator at $u=u_3$, we could first perform a global coordinate transformation $g$, then circle around the light operator, and then come back to the original coordinate system by transforming with $g^{-1}$. The physically meaningful information carried by the monodromy should not depend on the details defining the transformation $g$.   Note that a global coordinate transformation acts as
\be
P\xrightarrow{g} {{x P + y}\over{z P + t}}\, ,
\ee
for constants $x, y, z,$ and $t$. The inverse transformation is simply
\be
P\xrightarrow{g^{-1}} {{t P - y}\over{-z P +x}}\, .
\ee
With this we can figure out what the transformation $G^{-1}C_{\alpha}G$ looks like. One finds 
\be
P\xrightarrow{g^{-1}C_{\alpha}g} {{A' P +B'}\over{C' P +D'}}\, ,
\ee
where the constants $A',B',C',$ and $D'$ depend on both $A,B,C,D$ and the parameters defining the global transformation $g$. The fact that the new primed parameters depend on $g$ is equivalent to the statement that the parameters  $A,B,C,D$ are not gauge invariant and hence carry no physical meaning. One can go ahead and look for combinations of such parameters that do not depend on $g$. The two invariant objects are
\be
\begin{split}
I_1=\text{tr}M\, , \quad \text{and}\quad I_2=\text{tr}M^2\, ,
\end{split}
\ee
where we have defined the monodromy matrix
\be
M=\left(\begin{matrix}
   A &B\\
  C    & D 
\end{matrix}\right)\, ,
\ee
which is similar to the monodromy matrix appearing in the CFT$_2$ literature. We are now ready to write the gauge invariant information concerning the monodromy of $P$ for the case of the three point function. We find
\be
\begin{split}
I_1&=2\, , \\
\sqrt{{{I_1-I_2}\over 2}}& =2\pi \epsilon_{\alpha} \, .
\end{split}
\ee 
In the next section, we will compute these objects for the case of the four point function, where we circle around the pair of light operators. Demanding that the resulting monodromy matches the monodromy computed in this section will give rise to the BMS$_3$ block.

\subsection{Four-point function monodromy}\label{sec:Mmon4}
We turn now to the study of the correlator 
\be
\langle H(x_1) H(x_2) L (x_3) L(x_4) \rangle \, .
\ee
Just like before, we study the expectation value of the current $\mathcal{M}$. Using the BMS$_3$ Ward identities we have
\be
\begin{split}
{\cal M}(x_i;x)&={{\langle {\cal M}(x) H(x_1) H(x_2) L (x_3) L(x_4) \rangle }\over{\langle H(x_1) H(x_2)L (x_3) L(x_4)  \rangle }}   \\
&=\sum^4_{i=1} {{ \left({{\xi_i}\over{(u-u_i)^2}}+{1\over{u-u_i}}\partial_{v_i}\right) \langle H(x_1) H(x_2)L (x_3) L(x_4)  \rangle }\over{\langle H(x_1) H(x_2) L (x_3) L(x_4) \rangle }}\, .
\end{split}
\ee
if we define the auxiliary parameters 
\be
c_i={6\over{c_M}}\partial_{v_i} \log \langle {\cal M}(x) H(x_1) H(x_2)L (x_3) L(x_4)  \rangle\, ,
\ee
then we can write
\be
{\cal M}(x_i;x)=\sum^3_{i=1}  \left({{\xi_i}\over{(u-u_i)^2}}+{{c_M}\over 6}{{c_i}\over{u-u_i}}\right) \, .
\ee
The four-point function is not fixed by the symmetry of the field theory, and so the objects $c_i$ are unknown, unlike the case of the three-point function. However, they are constrained by the smoothness condition on the expectation value of the current. Recall the transformation law \ref{eq:Minf}. Under a global map $P(u)=1/u$, the current transforms as
\be
{\cal M}\rightarrow {\cal M}'(1/u)= u^4 {\cal M}(u) \, .
\ee
The condition that ${\cal M}'(0)$ is finite implies that ${\cal M}(u)\rightarrow u^{-4}$ as $u\rightarrow\infty$. This fixes three out of four objects $c_i$. For simplicity and without loss of generality we will place the operators at $u_1=0$, $u_2=\infty$, $u_3=1$, and we will leave $u_4=U$ as the free parameter (BMS$_3$ cross-ratio). In this case we have
\be
\begin{split}
c_1&=2\epsilon_L -U c_4\, ,\\
c_2&=0\, , \\
c_3&=-2\epsilon_{L}-(1-U)c_4\, .
\end{split}
\ee
The expectation value of the current is then
\be\label{eq:M4}
\begin{split}
{6\over{c_M}}{\cal M}(U;u)=&{1\over{u^2}}{{1-\beta^2}\over{4}}+\left({{1}\over{(u-U)^2}}+{{1}\over{u(1-u)^2}} +{{1}\over{u(1-u)}}  \right) \epsilon_L \\
&+\left( {1\over{U-u}}+{1\over u}-{U\over{u(1-u)}}   \right)c_4\, .
\end{split}
\ee
We now need to find the transformation $P(u)$ that takes us from the empty plane to a geometry whose expectation value of the current $\mathcal{M}$ matches \ref{eq:M4}. The differential equation to solve reads
 \be\label{eq:Peq4}
{1\over 2}S(P,u)={6\over{c_M}}{\cal M}(U;u) \, .
\ee
Note that to zeroth order in $\epsilon_L$ (and $c_4$), the differential equation matches the  one we solved for the case of the three-point function. We thus have the same solutions $g_i^{(0)}$. To linear order in  $\epsilon_L$, we again need to solve equation \ref{eq:diffO13}, but with ${\cal M}^{(1)}$ obtained from \ref{eq:M4}. The inhomogeneous equation is solved in the same fashion, and the monodromy of $P(u)$ can be computed again, this time circling around both the operators at $u_3=1$ and $u_4=U$. This means that when integrating the functions $f'_{i,j}(u)$, we pick up residues at both of these two points. After some algebra  which can be found in appendix \ref{app:Mmon}, the result reads
\be
\begin{split}
I_1&=2\, , \\
\sqrt{{{I_1-I_2}\over 2}}& =2\pi {{U^{-{{\beta}\over 2}}}\over{\beta}} \left[ c_4 U(U^{\beta}-1)+\epsilon_{L}\left( U^{\beta}(\beta+1)+\beta-1    \right)   \right]\, .
\end{split}
\ee
Comparing these results with the monodromy of the three point function, we obtain the following result for $c_4$
\be
c_4={{\beta U^{{{\beta}\over 2}}}\over{U(U^{\beta}-1)}}\epsilon_{\alpha} - {{U^{\beta}(\beta+1)+\beta-1}\over{U(U^{\beta}-1)}}\epsilon_L\, .
\ee
We conclude that the BMS$_3$ block has the following form
\be\label{eq:FM}
{\cal F}_{\alpha}= F(U) e^{V\left(  {{\beta U^{{{\beta}\over 2}}}\over{U(U^{\beta}-1)}}\xi_{\alpha} - {{U^{\beta}(\beta+1)+\beta-1}\over{U(U^{\beta}-1)}}\xi_L   \right)}\, ,
\ee
where $F(U)$ is some undetermined function of $U$, which we must obtain by studying the monodromy concerning the current ${\cal L}$. We turn now to this issue.

\subsection{The monodromy of ${\cal L}$ - Three-point function}
In the last subsections we have been able to obtain only part of the BMS$_3$ block. The reason for this is that the monodromy method only computes derivatives of the conformal block with respect to the cross ratios, and we have only obtained one of such derivatives. In this section we obtain the rest of the block by applying the philosophy of the geometric monodromy method to the expectation value of the current ${\cal L}$. The computations in this sections are a bit more complicated due to the way the current ${\cal L}$ transforms under finite BMS$_3$ transformations. Just like we did with the other current, we start by considering the three-point correlator  
\be
\langle H(x_1) H(x_2) L_{\alpha} (x_{\alpha}) \rangle \, ,
\ee
and inserting the current ${\cal L}$ to compute an expectation value. Using the BMS$_3$ Ward identities we have
\be
{\cal L}(x,x_i)={{ \sum^n_{i=1} \left({{\Delta_i}\over{(u-u_i)^2}}-{1\over{u-u_i}}\partial_{u_i}+2\xi_i{{(v-v_i)}\over{(u-u_i)^3}}+{{v-v_i}\over{(u-u_i)^2}}\partial_{v_i}\right)\langle H(x_1) H(x_2) L_{\alpha} (x_{\alpha}) \rangle}\over{\langle H(x_1) H(x_2) L_{\alpha} (x_{\alpha}) \rangle}} \, .
\ee
We now use our previous definition of the objects $c_i$, which we recall here
\be
c_i={6\over{c_M}}\partial_{v_i} \log \langle {\cal M}(x) H(x_1) H(x_2) L_{\alpha} (x_{\alpha}) \rangle\, ,
\ee
and introduce four new objects;
\be
d_i={6\over{c_M}}\partial_{u_i} \log \langle {\cal M}(x) H(x_1) H(x_2) L_{\alpha} (x_{\alpha}) \rangle\, .
\ee
With these definitions at hand we can re-write the current as
\be
{6\over{c_M}}{\cal L}(x,x_i)= \sum^n_{i=1} \left({{\delta_i}\over{(u-u_i)^2}}-{1\over{u-u_i}}d_i+2\epsilon_i{{(v-v_i)}\over{(u-u_i)^3}}+{{v-v_i}\over{(u-u_i)^2}}c_i\right) \, .
\ee
where we have also introduced $\delta_i=6\Delta_i/c_M$. Using an explicit expression for the invariant BMS$_3$ three-point function we can obtain the objects $c_i$ and $d_i$. We can also use a coordinate transformation to move the heavy operators to $u_1=v_1=0$, $u_2=\infty$, and $v_2=0$. We leave the third operator at an arbitrary location in the null plane. This results in the following simple expression for our current
\be
{6\over{c_M}}{\cal L}(x,x_i)= {{\delta_H}\over{u^2}} +2{{v}\over{u^3}}\epsilon_H+{{u_3}\over{u(u-u_3)^2}}\delta_L +{{v u_3 (3u-u_3)-u v_3(u+u_3)}\over{u^2(u-u_3)^3}}\epsilon_L\, .
\ee
we now would like to solve the following differential equation for the coordinate transformation $G$
\be
\begin{split}
{6\over{c_M}}{\cal L}(x,x_i)&= {{1}\over{2}}{{G'\left(P'P''-3\left(P''\right)^3\right)+P'\left(3P''G''-P'G'''\right)}\over{\left(P'\right)^3}}\\
&-{{1}\over{2}}v{{3\left(P''\right)^2-4P'P''P'''+\left(P'\right)^2P'''}\over{\left(P'\right)^3}}\, .
\end{split}
\ee
Note that the function $P$ has been studied in previous subsections, and it is responsible for turning on the appropriate current ${\cal M}$, so this differential equation is exclusively for the function $G$. Indeed, note that ${\cal M}$ is related to the Schwarzian derivative in such a way that
\be
v\left(  {6\over{c_M}}{\cal M}(u_i;u) \right)^{\prime}={{1}\over{2}}v{{3\left(P''\right)^2-4P'P''P'''+\left(P'\right)^2P'''}\over{\left(P'\right)^3}}\, .
\ee
and so instead of working with ${\cal L}$ on its own, we will study the combination
\be
{6\over{c_M}}\tilde{\cal L}(u_i;u)={6\over{c_M}}\left({\cal L}(u_i;u)+v\left( {\cal M}(u_i;u) \right)^{\prime}\right)= {{\delta_H}\over{u^2}} +{{u_3}\over{(u-u_3)^2}}\delta_{\alpha} -{{u v_3(u+u_3)}\over{u(u-u_3)^3}}\epsilon_{\alpha}\, ,
\ee
such that the differential equation to solve does not depend on the coordinate $v$, and reads
\be
{6\over{c_M}}\tilde{\cal L}(x,x_i)= {{1}\over{2}}{{G'\left(P'P''-3\left(P''\right)^3\right)+P'\left(3P''G''-P'G'''\right)}\over{\left(P'\right)^3}}\, .
\ee
A simplification can be achieved through the replacement $G(u)=P'(u)X(u)$, which leads to
\be\label{eq:diffeqX}
{6\over{c_M}}\tilde{\cal L}=-{1\over 2}X'''-2X' {6\over{c_M}}{\cal M} -X {6\over{c_M}}{\cal M}'\, .
\ee
We now need to solve this differential equation order by order in the quantum numbers of the light operators. Once this is done, we will study the monodromy of the family of solutions when we cycle around the insertion of the light primary operator. The details concerning this computation can be found in appendix \ref{app:MONODROMYL3}. As when studying the monodromy of ${\cal M}$, not all the components of the monodromy are gauge invariant. The gauge invariant information can be packaged in an object that we call $I_3$. The result derived in appendix \ref{app:MONODROMYL3} reads
\be\label{eq:i33p}
\sqrt{I_3}=4\pi \sqrt{{{\delta_{\alpha}\epsilon_{\alpha}}\over{\beta}}}\, .
\ee
The objective of next section is to compute this invariant in the case of the four point function, and compare the result to equation \ref{eq:i33p}.

\subsection{The monodromy of $\mathcal{L}$ - Four-point function}
We turn our attention to the four-point function of two heavy and two light operators,
\be
\langle H(x_1) H(x_2) L (x_3) L(x_4) \rangle \, .
\ee
We study the expectation value of $\mathcal{L}$, and we look for the gauge transformation that would take us from the null plane to a geometry whose expectation value for the current matches the one in the presence of the four operators in question. Making use of the BMS$_3$ Ward identities we arrive at
\be
{6\over{c_M}}{\cal L}(x,x_i)= \sum^n_{i=1} \left({{\delta_i}\over{(u-u_i)^2}}-{1\over{u-u_i}}d_i+2\epsilon_i{{(v-v_i)}\over{(u-u_i)^3}}+{{v-v_i}\over{(u-u_i)^2}}c_i\right) \, ,
\ee
where $c_i$ and $d_i$ are related to the derivatives of the block with respect to $v_i$ and $u_i$ respectively. Smoothness at infinity implies ${\cal L}(u)\rightarrow u^{-4}$ as $u\rightarrow\infty$. This fixes all $d_i$ except one. Without loss of generality, we will place the operators at $u_1=0$, $u_2=\infty$, $u_3=1$, $u_4=U$, $v_1=v_2=v_3=0$, and $v_4=V$. This leads to the following expression for the current expectation value
\be\label{eq:Ltilde4}
\begin{split}
{6\over{c_M}}\tilde{\cal L}(x,x_i)=&\frac{\delta _{H}}{u^{2}}+\left[ \frac{1}{\left(
u-U\right) ^{2}}+\frac{2-u}{u\left( u-1\right) ^{2}}\right] \delta _{L}+%
\frac{U\left( U-1\right) }{u\left( u-1\right) \left( u-U\right) }d_{4}\\
&+V\left[ \frac{2}{\left( U-u\right) ^{3}}\epsilon_{L}+\frac{1}{u\left(
u-1\right) }c_{4}\right]\, .
\end{split}
\ee
The differential equation to solve is
\be
{6\over{c_M}}\tilde{\cal L}(x,x_i)= {{1}\over{2}}{{G'\left(P'P''-3\left(P''\right)^3\right)+P'\left(3P''G''-P'G'''\right)}\over{\left(P'\right)^3}}\, ,
\ee
Under the replacement $G(u)=P'(u)X(u)$ we obtain the differential equation \ref{eq:diffeqX}. In appendix \ref{app:MONODROMYL4}, we solve this differential equation to linear oder in the light operators' quantum numbers. The invariant $I_3$ introduced in the previous section can then be extracted. Comparing with formula \ref{eq:i33p} we arrive at the following expression for the auxiliary parameter 
\be\label{eq:d4}
\begin{split}
d_{4} &=\frac{\beta U^{\frac{\beta }{2}}}{U\left( U^{\beta }-1\right) }%
\delta _{I}-\frac{\left( U^{\beta }\left( \beta +1\right) +\beta
-1\right) }{U\left( U^{\beta }-1\right) }\delta _{M} \\
&+\left[ V\frac{1-\beta +U^{\beta }\left( \beta +1\right) \left(
U^{\beta }+2\left( \beta -1\right) \right) }{U^{2}\left( U^{\beta
}-1\right) ^{2}}+\delta _{H}\frac{2-2U^{2\beta }+4U^{\beta }\beta \log U}{%
\beta U\left( U^{\beta }-1\right) ^{2}}\right] \epsilon _{M} \\
&+\left[ V\frac{U^{\frac{\beta }{2}}\beta \left( 2-\beta -U^{\beta
}\left( \beta +2\right) \right) }{2U^{2}\left( U^{\beta }-1\right) ^{2}}%
+\delta _{H}\frac{U^{\frac{\beta }{2}}\left( 2\left( U^{\beta }-1\right)
-\left( U^{\beta }+1\right) \beta \log U\right) }{\beta U\left( U^{\beta
}-1\right) ^{2}}\right] \epsilon _{I} \, .
\end{split}
\ee
It is reassuring that the following is true
\be
\frac{d}{dV}d_{4}=\frac{d}{dU}c_{4}\, .
\ee
Integrating $d_{4}$ with respect to $U$ yields the following answer for the BMS$_3$ block
\be\label{eq:ResultBMSFT}
{\cal F}_{\alpha} = \left( {{1-U^{{\beta}\over{2}}}\over{1+U^{{\beta}\over{2}}}}  \right)^{\Delta_{\alpha}}   \left( {{   U^{\beta-1}    }\over{  (1-U^{\beta})^2    }}   \right)^{\Delta_L} e^{V\left(  {{\beta U^{{{\beta}\over 2}}}\over{U(U^{\beta}-1)}}\xi_{\alpha} - {{U^{\beta}(\beta+1)+\beta-1}\over{U(U^{\beta}-1)}}\xi_L \right)
+\Delta_H\left(     {{2U^{{\beta}\over 2}}\over{\beta (U^{\beta}-1)}}    \xi_{\alpha}            +  {{2(U^{\beta}+1)}\over{\beta(1-U^{\beta})}} \xi_L                    \right)  \log U 
}\, .
\ee
This is the main result of this paper. Note that this also agrees with the $V$-dependent part found in \ref{eq:FM}. 

It is interesting to check that the result found here is invariant under an imaginary shift of the coordinates $u,\phi$, which are related to $U,V$ through the map \ref{eq:map}. In order to keep it simple, we can analyze the result \ref{eq:ResultBMSFT} in the case of the vacuum block ($\xi_{\alpha}=0$). It is then easy to check that the result is invariant under the following shift
\be
\begin{split}
u&\rightarrow u+\pi i {J\over M}\, ,\\
\phi &\rightarrow \phi +2\pi i {1\over {\sqrt{M}}}\, ,
\end{split}
\ee
where the parameters $M, J$ are related to the quantum numbers of the heavy operators as in equation \ref{eq:MJheavy}. This is an indication that correlators of light operators in presence of highly energetic operators exhibit thermal behavior in a BMSFT. One of the objectives of this paper is to show how these correlators can be computed holographically in a thermal background. This will be the focus of section \ref{sec:Holo}.

This concludes the computation of the BMS$_3$ block in the field theory side, which relies only on the structure of the BMS$_3$ algebra. The approach presented here is not the usual monodromy method used in conformal field theories to compute Virasoro and global blocks. Usually, the differential equations giving rise to a monodromy problem are obtained from the study of null vectors of the symmetry algebra. This will be the topic of section \ref{sec:null}, where we will first review  some of the CFT calculations in the literature, and then study the null vectors of the BMS$_3$ algebra to obtain the same results we have obtained through the geometric monodromy approach.

\section{Monodromy method from BMS$_3$ null vectors}\label{sec:null}
In the previous section we have computed BMS$_3$ blocks by constraining the monodromy of the  coordinate transformations that turn on the appropriate currents. In this section we will re-derive the differential equations involved in the monodromy method by studying null vectors of the BMS$_3$ algebra. The construction presented here is very different from the one used in two-dimensional conformal field theories. The main difference is that the BMS$_3$ null vectors built out of descendants of a particular primary are not light, and as a consequence cannot be used to obtain a useful differential equation. This will lead to the introduction of BMS$_3$ multiplets, which are reducible but indecomposable representations of the BMS$_3$ algebra, much like the ones studied in the context of logarithmic CFT's.  We start with a quick review of the monodromy method in CFT$_2$. 

\subsection{CFT$_2$ null vectors}
The study of degenerate representations of the Virasoro algebra was first presented in \cite{Belavin:1984vu}. The techniques developed in that paper were used in \cite{Fitzpatrick:2014vua} to compute the Virasoro blocks of a four-point function involving two heavy and two light operators. In this section we also follow closely \cite{Harlow:2011ny,Hartman:2013mia}. One of the insights presented in \cite{Belavin:1984vu} is that the representations of the Virasoro algebra are irreducible unless the conformal dimension $h$ labeling the representation takes some special values. For these values, there exists a special null vector $\vert \chi \rangle$ satisfying the following properties
\be\label{eq:nulldef}
\begin{split}
L_n \vert \chi \rangle &=0\, , \quad \, \text{for} \quad n>0\, , \\
L_0 \vert \chi \rangle &= (h+K)\vert \chi \rangle\, ,
\end{split}
\ee
for a positive integer $K$. For example, for $K=2$, we have the following null vector
\be\label{eq:nullcft}
\vert \chi \rangle =\left(  L_{-1}^2-{{2(1+2h)}\over{3}}L_{-2}  \right)\vert h \rangle\, .
\ee
Here $\vert h\rangle$ is the highest weight state of conformal dimension $h$. The vector in \ref{eq:nullcft} is null for the following choice of conformal dimension
\be
h_{\pm}={{5-c\pm \sqrt{(c-1)(c-25)}}\over{16}}\, .
\ee
In the large central charge limit, the conformal dimension $h$ with positive sign choice is ${\cal O}(1)$, rendering the highest weight state $\vert h\rangle$ light. Explicitly, the conformal dimension of this light state reads
\be
h\rightarrow h_+ =-{1\over 2}-{9\over {2c}}+{\cal O}(c)^{-2}\, .
\ee
Note that equations \ref{eq:nulldef} imply that the vector $\vert \chi \rangle$ can be considered the highest weight state of its own Verma modulus with conformal dimension $h+K$. However, this would imply that the representation labeled by $h$ is reducible. We thus need to formally set the null vector to zero
\be\label{eq:shorteningCFT}
\vert \chi \rangle  = \left(  L_{-1}^2-{{2(1+2h)}\over{3}}L_{-2}  \right)\vert h \rangle =0 \, .
\ee
This equation can be used to compute Virasoro blocks. The strategy is to insert the light operator  ${\cal O}_{h_+}$ associated to the state $\vert h_+ \rangle$ in a CFT four-point function, and then apply the Virasoro generators in \ref{eq:nullcft} to obtain a differential equation for the resulting five-point function. The fact that the operator is light implies that the leading semi-classical behavior of the correlator is left unchanged. More concretely, we have
\be
\langle {\cal O}_1 {\cal O}_2 {\cal O}_{h_+}(z) {\cal O}_3 {\cal O}_4 \rangle = \Psi (z;z_i)  \langle {\cal O}_1 {\cal O}_2 {\cal O}_3 {\cal O}_4 \rangle\, ,
\ee
where the wave function $\Psi (z;z_i)$ introduced here (and its derivatives) are ${\cal O}(e^{c^0})$. Applying the shortening condition \ref{eq:shorteningCFT} and using the Ward identities of the Virasoro algebra leads to the following differential equation
\be
\Psi'' (z;z_i)+{6\over{c}}T (z;z_i) \Psi (z;z_i)=0\, .
\ee
Studying the solution space of this differential equation in the same fashion as in section \ref{sec:BMSFT} leads to explicit expressions for the Virasoro blocks of the correlator $\langle {\cal O}_1 {\cal O}_2 {\cal O}_3 {\cal O}_4 \rangle$. We now try to apply this logic in field theories with BMS$_3$ symmetry.

\subsection{BMS$_3$ null vectors}
Given the success of the study of CFT$_2$ null vectors, it seems worth analyzing the existence of such objects in field theories with BMS$_3$ symmetry. As we will see below, null vectors constructed out of irreducible representations of the Virasoro algebra will not be useful to obtain a differential equation with non-trivial monodromy. Some of the statements made in this section can also be found in \cite{Bagchi:2009pe}. We look for vectors obeying the following conditions
\be
\begin{split}
L_0 \vert \chi \rangle &= \Delta' \vert \chi \rangle\, ,\quad \, M_0 \vert \chi \rangle = \xi' \vert \chi \rangle\, ,\\
L_n  \vert \chi \rangle &= M_n\vert \chi \rangle=0 \, ,\quad \text{for} \quad n>0\, .
\end{split}
\ee
There are two possible null vectors obeying these conditions and so we must formally set them to zero. They read
\be
\begin{split}
\vert \chi_1 \rangle &= M_{-1}^2 \vert 0,\Delta \rangle =0\, , \\
\vert \chi_2 \rangle &= \left( M_{-1}^2 +{{c_M}\over 6} M_{-2} \right) \vert -{{c_M}\over 8} , \Delta \rangle=0 \, .
\end{split}
\ee
Here we have used the notation $\vert \xi, \Delta\rangle$ to represent the highest weight state in a BMS representation with $\xi$ and $\Delta$ as the eigenvalues of $M_0$ and $L_0$ respectively. The first null vector is constructed from a light BMS primary. This implies that it can be inserted into a correlator without affecting the semi-classical structure. However, the differential equation implied by the shortening condition is trivial. It simply indicates that the resulting wave function is linear in the coordinate $v$, which does not give non-trivial information about the monodromy of the solutions around the location of operators in the null plane. The second null vector involves a heavy BMS primary in the sense that the quantum number $\xi$ scales with $c_M$. This implies that inserting this primary into a BMS correlator changes the semi-classical structure. We conclude that BMS$_3$ null vectors cannot be used to obtain a differential equation with non-trivial monodromy.

\subsection{BMS$_3$ multiplets}\label{sec:multiplets}
So far we have  only tried constructing a useful null vector out of irreducible representations of the symmetry algebra. It turns out the problem can be solved by analyzing reducible but indecomposable representations of the BMS$_3$ algebra, that we will dub as BMS$_3$ multiplets. These representations are characterized by a non-diagonal action of the center of the algebra on a set of operators. These representations are the main ingredient in logarithmic conformal field theories (see \cite{Hogervorst:2016itc} for a modern review). logCFT's can be used to study many interesting models, like percolation \cite{Cardy:1999zp} and systems with quenched disorder \cite{Caux:1995nm} among many others. In this subsection we will generalize logarithmic multiplets to field theories with BMS symmetry. It is worth pointing out that the representations presented in this section need not be part of the physical content of the field theory. The same way the light primary operator used to construct a CFT$_2$ null vector is a purely mathematical tool, we will also regard BMS multiplets as such.

A rank-$r$ multiplet consists of $r$ operators such that the action of $L_0$ and $M_0$ adopts a Jordan block form. In this section we will study rank $2$  multiplets, such that
\be
\begin{split}
L_0 \vert \tilde{\psi}_1 \rangle &= \Delta_1 \vert \tilde{\psi}_1 \rangle \, , \\
L_0 \vert \tilde{\psi}_2 \rangle &= \Delta_2  \vert \tilde{\psi}_2 \rangle+ \left( L_0 \right)_{21}  \vert \tilde{\psi}_1 \rangle \, ,
\end{split}
\ee
and 
\be
\begin{split}
M_0 \vert \tilde{\psi}_1 \rangle &= \xi_1 \vert \tilde{\psi}_1 \rangle \, , \\
M_0 \vert \tilde{\psi}_2 \rangle &= \xi_2  \vert \tilde{\psi}_2 \rangle+ \left( M_0 \right)_{21}  \vert \tilde{\psi}_1 \rangle \, .
\end{split}
\ee
We also demand that all positive modes of the BMS algebra annihilate the stares $\vert \tilde{\psi}_i \rangle$. We are hoping to construct a new rank 2 multiplet by acting on the operators $\tilde{\psi}_1 $ and $\tilde{\psi}_2$ with the negative modes of the BMS$_3$ algebra. The existence of such multiplet would render the original BMS$_3$ multiplet spanned by $\vert \tilde{\psi}_1 \rangle$ and $\vert \tilde{\psi}_2 \rangle$ not only reducible but decomposable. We thus formally set the new multiplet to zero. This gives rise to a null BMS$_3$ multiplet we can hope to use to obtain  differential equations with non-trivial monodromy structure. Similar kinds of degenerate multiplets have been studied in the context of logarithmic CFT's in \cite{Flohr:1997wm,Flohr:2000mc}. 

Much like the case of CFT null vectors, we do not expect a degenerate multiplet to exist for any value of the quantum numbers and matrix elements of $L_0$ and $M_0$ defining the original multiplet. Performing some algebra, we find that the right choice is
\be\label{eq:OriginalMultiplet}
\begin{split}
L_0 \vert \tilde{\psi}_1 \rangle &= -{1\over 2}\vert \tilde{\psi}_1 \rangle \, , \\
L_0 \vert \tilde{\psi}_2 \rangle &= -{1\over 2}  \vert \tilde{\psi}_2 \rangle-{9\over {c_M}}  \vert \tilde{\psi}_1 \rangle \, , \\
M_0 \vert \tilde{\psi}_1 \rangle &=0 \, , \\
M_0 \vert \tilde{\psi}_2 \rangle &= -{1\over 2}  \vert \tilde{\psi}_1 \rangle \, .
\end{split}
\ee
Note that all the matrix elements of $L_0$ and $M_0$ do not diverge in the large $c_M$ limit. This means that the operators $\tilde{\psi}_1$ and $\tilde{\psi}_2$ can be inserted into BMS correlators without changing the semi-classical structure. To be more concrete, we will write 
\be
\langle {\cal O}_1 {\cal O}_2 \tilde{\psi}_i {\cal O}_3 {\cal O}_4 \rangle =\psi_i (u,v;u_i,v_i) \langle {\cal O}_1 {\cal O}_2 {\cal O}_3 {\cal O}_4 \rangle\, ,
\ee
where the wave functions $\psi_i$ introduced here (and its derivatives) are ${\cal O}(e^{c_M^0})$.

 For the choice of operators in \ref{eq:OriginalMultiplet}, there are two level one null vectors which read
\be
\begin{split}
\vert \chi^{(1)}_1\rangle &=M_{-1} \vert \tilde{\psi}_1 \rangle=0\, ,\\
\vert \chi^{(1)}_2\rangle &=M_{-1} \vert \tilde{\psi}_2 \rangle-L_{-1}\vert \tilde{\psi}_1 \rangle=0\, .
\end{split}
\ee
At level two, we find the following null vectors
\be
\begin{split}
\vert \chi^{(2)}_1\rangle &=\left(  L_{-1}^2 +{6\over{c_M}}M_{-2}   \right) \vert \tilde{\psi}_1 \rangle=0\, ,\\
\vert \chi^{(2)}_2\rangle &= \left(  L_{-1}^2 +{6\over{c_M}}M_{-2}   \right)     \vert \tilde{\psi}_2 \rangle         +{6\over{c_M}} \left(   {{26}\over{c_M}} M_{-2} +L_{-2}     \right)                       \vert \tilde{\psi}_1 \rangle=0\, .
\end{split}
\ee
Inserting the null vectors into BMS correlators yields the following differential equations for the wave functions $\psi_i$
\be
\begin{split}
\partial_v \psi_1 &=0\, , \\
\partial_v \psi_2 +\partial_u \psi_1 &=0\, , \\
\left( \partial_u^2+{6\over{c_M}} {\cal M} \right) \psi_1 &=0\, , \\
\left( \partial_u^2+{6\over{c_M}} {\cal M} \right) \psi_2+{6\over {c_M}} \left(  {{26}\over{c_M}}{\cal M} +{\cal L}    \right)\psi_1&=0\, .
\end{split}
\ee
The first two equations simply imply $\psi_1=\psi_1(u)$ and $\psi_2=\psi_2(u)-v \partial_u \psi_1$. Inserting these expressions into the other two differential equations, simplifying, and taking the large $c_M$ limit leads to
\be
\begin{split}
\left( \partial_u^2+{6\over{c_M}} {\cal M} \right) \psi_1(u) &=0\, , \\
\left( \partial_u^2+{6\over{c_M}} {\cal M} \right) \psi_2+{6\over {c_M}} \left( {\cal L}+v \partial_u{\cal M}    \right)\psi_1&=0\, .
\end{split}
\ee
Note that the first equation here is equivalent to formula \ref{eq:diffg} used in previous sections to partially compute BMS blocks. The second equation can be further manipulated into a more familiar form. We change variables as $\psi_2=-\partial_u \psi_1 X+\psi_1 \partial_u X/2$, which leads to
\be
{6\over{c_M}}\tilde{\cal L}=-{1\over 2}X'''-2 X' {6\over{c_M}}{\cal M}-X {6\over{c_M}}{\cal M}'\, .
\ee
Here, we have introduced $\tilde{\cal L}={\cal L}+v \partial_u{\cal M}$. This matches the differential equation \ref{eq:diffeqX}, which was used to complete the calculation of the block. This finalizes  our analysis of the Monodromy method in the field theory. We have shown that studying light degenerate BMS multiplets is equivalent to the geometric monodromy method presented in previous sections.

We turn now to the holographic interpretation of BMS blocks. In section \ref{sec:Holo} we will find that the blocks we have computed in the field theory can be obtained by considering probe particles propagating in a flat space cosmology.


\section{Holographic BMS$_3$ blocks}\label{sec:Holo}
In this section we explore the holographic computations leading to formula \ref{eq:ResultBMSFT} for the BMS$_3$ block.  In previous work \cite{Hijano:2017eii} we presented the holographic picture of Poincar\'e blocks (BMS global blocks) which had been computed in the field theory in \cite{Bagchi:2017cpu}. The picture was inspired by a holographic computation of entanglement entropy in two dimensional BMSFT's \cite{Jiang:2017ecm}. The construction is in some ways similar to the holographic picture found in the AdS/CFT context in \cite{Hijano:2015rla,Hijano:2015zsa}. They key ingredient in flat space is the introduction of a new extrapolate dictionary. It consists on integrating position space Feynman diagrams along the null geodesics falling from the null boundary at the location of the BMS primary operators. 
\begin{align}\label{eq:Extrapolate}
\langle \cO_1(x_1) \cO_2(x_2) \ldots \rangle 
\sim \int_{\gamma_{x_1}} d\lambda_1 \int_{\gamma_{x_2}} d\lambda_2 \ldots 
\langle \Psi_1(\lambda_1) \Psi_2(\lambda_2) \ldots \rangle\, .
\end{align}
 Here $\Psi_i$ are bulk fields propagating in Minkowski space-time, and the integrals run along the null geodesics $\gamma_{x_i}$, which fall from the null boundary at the locations $x_i$ where the BMS field theory local primary operators are placed. If the bulk fields are sufficiently massive, the integrals inside the bulk Feynman diagram and the integrals over the null geodesics can be well approximated by a saddle point approximation. The computation of the correlators can then be replaced by an optimization problem, where one extremizes the length of a geodesic network attached to the null lines $\gamma_{x_i}$. 

The novel feature in this work is the fact that we are now considering two of the operators to be heavy enough to backreact the background geometry non-trivially. Einstein gravity in three dimensions and in absence of a cosmological constant has a family of solutions known as flat space cosmological solutions (FSC). The line element reads
\be\label{eq:ds}
ds^2=M du^2 -2 du dr +J du d\phi +r^2 d\phi^2\, .
\ee
Here the parameters $M$ and $J$ parametrise the ADM mass and angular momentum of the gravitational solution. These cosmologies were first considered in string theory as  shifted-boost orbifolds of Minkowski spacetimes \cite{Cornalba:2002eg,Cornalba:2003kd}. Note that the case $M=-1$, $J=0$ corresponds to Global Minkowski, whose holographic dual is the empty cylinder BMSFT. In order for the solution to be understood as a cosmology, it is required that $M>0$. We expect that the parameters $M$ and $J$ can be related to the quantum numbers $\xi_H$ and $\Delta_H$ of the heavy operators participating in the four-point function studied in previous sections.  The line element \ref{eq:ds} can be written in ADM form \cite{Barnich:2012aw}
\be\label{eq:ADMform}
ds^2=-\left( {{J^2}\over{4 \tilde{r}^2}}  -M \right) d\tilde{t}^2  +{1\over{ {{J^2}\over{4\tilde{r}^2}}  -M   }} d\tilde{r}^2 +\tilde{r}^2 \left(  d\tilde{\phi}+{{J}\over{2\tilde{r}^2}}d\tilde{t}     \right)^2\, .
\ee
There is a special null Cauchy Horizon at
\be
\tilde{r}=R_c\, , \quad \text{with}\quad  R_c={{J}\over{2\sqrt{M}}}\, .
\ee
This hypersurface is also a Killing Horizon, with a Hawking temperature associated to it. This means that the bulk state associated to this background geometry is in a sense thermal. 

The light operators are expected to behave as bulk probes. As we have already mentioned, in previous work we computed holographic correlators involving exclusively light operators. Correlators were computed by attaching geodesic networks to the null lines falling from the boundary at the locations of the light primaries. Extremizing the length of such networks yields the correct expression for Poincar\'e blocks. The new approach is to introduce geodesic networks connecting  the horizon of the FSC \ref{eq:ds} to the null lines falling from the boundary at the location of the light primaries. Extremizing the length of such objects will yield expression \ref{eq:ResultBMSFT}. As a warm up, we will fist study the BMS$_3$ block involving spinless primaries.

\subsection{Spinless holographic BMS$_3$ blocks}
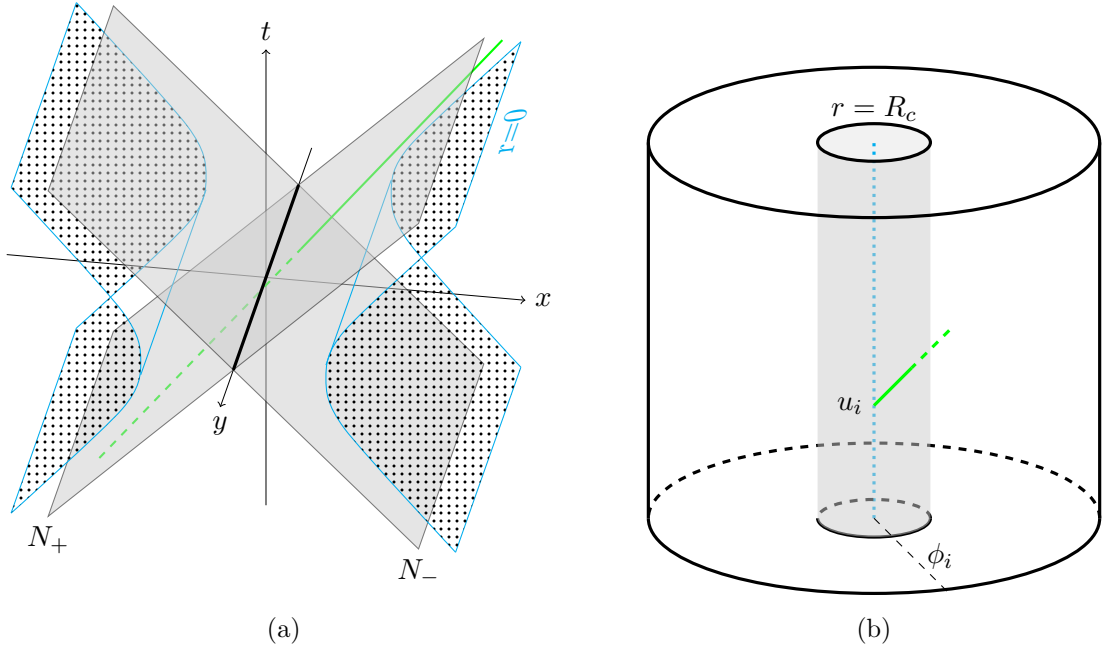
\begin{figure}[t!]
\centering
\begin{subfigure}[t]{0.48\textwidth}
        \centering

\tdplotsetmaincoords{60}{100}
\begin{tikzpicture}[scale=5,tdplot_main_coords]

\draw [->,black,line width=0.1mm] (1/2,1/2,-0.2)--(1/2,1/2,1.2 ) node[pos=1,above]{$t$};
\draw [->,black,line width=0.1mm] (1/2,-0.2,0.5)--(1/2,1.2,0.5) node[pos=1,right]{$x$};
\draw [->,black,line width=0.1mm] (-0.2,1/2,0.5)--(1.2,1/2,0.5) node[pos=1,below]{$y$};

    \def\x{1}

  \filldraw[
        draw=black,%
        fill=black!20,
         opacity=0.5,%
    ]          (0,1,0)
            -- (1,1,0)
            -- (1,1/2,1/2)
            -- (0,1/2,1/2)
            -- cycle;

\draw [green, thick] (0,1+0.05,1)--(1/4,1/2+0.05,1/2-0.05  );
\draw [green, thick,dashed] ( 1/2,0.05,-0.1      )--(1/4,1/2+0.05,1/2-0.05  );


\draw [cyan,name path=Lpath] plot [smooth, tension=0.5] coordinates { (1,1+0.1,0) (1,1/2+0.3,1/2-0.15)  (1,1/2+0.25,1/2) (1,1/2+0.3,1/2+0.15) (1,1+0.1,1)};
\draw [cyan] (1,1+0.1,0)--(0,1+0.1,0);
\draw [cyan,name path=Rpath] plot [smooth, tension=0.5] coordinates { (0,1+0.1,0) (0,1/2+0.3,1/2-0.15)  (0,1/2+0.25,1/2) (0,1/2+0.3,1/2+0.15) (0,1+0.1,1)};
\draw [cyan,name path=Upath] (1,1+0.1,1)--(0,1+0.1,1)  node[midway,sloped,below,xslant=-0.5] {r=0};
\tikzfillbetween[
    of=Lpath and Rpath,split
  ] {pattern=dots};

\draw [cyan,name path=Mpath] (0,1/2+0.25,1/2+0.04)--(1,1/2+0.25,1/2+0.04);

    \filldraw[
        draw=black,%
        fill=black!20,
        opacity=0.5,%
    ]          (0,0,0)
            -- (1,0,0)
            -- (1,1,1)
            -- (0,1,1)
            -- cycle;

\draw [cyan,name path=Lpath2] plot [smooth, tension=0.5] coordinates { (1,0-0.1,0) (1,1/2-0.3,1/2-0.15)  (1,1/2-0.25,1/2) (1,1/2-0.3,1/2+0.15) (1,0-0.1,1)};
\draw [cyan] (1,0-0.1,0)--(0,0-0.1,0);
\draw [cyan,name path=Rpath2] plot [smooth, tension=0.5] coordinates { (0,0-0.1,0) (0,1/2-0.3,1/2-0.15)  (0,1/2-0.25,1/2) (0,1/2-0.3,1/2+0.15) (0,0-0.1,1)};
\draw [cyan,name path=Upath] (1,0-0.1,1)--(0,0-0.1,1);

\tikzfillbetween[
    of=Lpath2 and Rpath2,split
  ] {pattern=dots};
\draw [cyan,name path=Mpath] (0,1/2-0.25,1/2-0.04)--(1,1/2-0.25,1/2-0.04);


  \filldraw[
        draw=black,%
        fill=black!20,
         opacity=0.5,%
    ]          (0,0,1)
            -- (1,0,1)
            -- (1,1/2,1/2)
            -- (0,1/2,1/2)
            -- cycle;

\draw (1,0,0) node[below] {\textcolor{black}{$N_+$}};
\draw (1,1,0) node[below] {\textcolor{black}{$N_-$}};

  \draw[
       black,very thick%
    ]          (0,1/2,1/2)
            -- (1,1/2,1/2);

\end{tikzpicture}

        \caption{ }
    \end{subfigure}%
\quad
\begin{subfigure}[t]{0.48\textwidth}
        \centering

\begin{tikzpicture}

 \fill[color=black!20, opacity=0.25](2,7) ellipse (0.75 and 0.25);

\draw[white,very thick,name path=TOP](1.25,7) arc (180:360:0.75 and 0.25);

\draw[cyan,very thick,dotted] (2,2) -- (2,7) node [pos=1,above=2]{{$ $}} node [pos=0,below=2]{{$ $}};




\draw[black,very thick,dashed](2.75,2) arc (0:180:0.75 and 0.25);
\draw[black,very thick,name path=BOT](1.25,2) arc (180:360:0.75 and 0.25);

\draw[black,very thick,dashed](5,2) arc (0:180:3 and 1);
\draw[black,very thick](-1,2) arc (180:360:3 and 1);




 \fill[color=black!20, opacity=0.5]
(1.25,7)--(1.25,2)
-- (1.25,2) arc (180:360:0.75 and 0.25)
-- (2.75,7) 
-- (2.75,7) arc (360:180:0.75 and 0.25) --cycle;


\draw[green,very thick] (2,3.5) -- (2.5,4) node[pos=0,left]{\textcolor{black}{$u_{i}$}};
\draw[green,very thick,dashed] (3,4.5) -- (2.5,4);

\draw[black,very thick] (-1,2) -- (-1,7);
\draw[black,very thick] (5,2) -- (5,7);

\draw[black,very thick](2,7) ellipse (0.75 and 0.25);

\draw[black,very thick](2,7) ellipse (3 and 1)  node [above=4] {$r=R_c$};

\draw[black,dashed](2,2) --(3,1) node [pos=0.5,right=2] {$\phi_i$};

\end{tikzpicture}


        \caption{ }
    \end{subfigure}
    \caption{a) Minkowski Space coordinates. The gray planes correspond to the Cauchy horizon. The planes have been spelled out in equation \ref{eq:FSCMINK}. They intersect at the special geodesic $\gamma_{\cal H}$ placed at $t=x=0$. The dotted surface corresponds to the origin in ADM coordinates. The green line represents a null geodesic. The dashed part is time-like separated from the Cauchy Horizon.      b) ADM coordinates. The gray cylinder is the Cauchy Horizon at $r=R_c$. The dotted blue line is the origin $r=0$.  The green line is a null geodesic that intersects the null boundary at $u_i,\phi_i$. }\label{fig:setup}
\end{figure}

Flat space cosmologies can be thought of as quotients of empty flat space. This is a feature of gravity in three dimensions, and it is also true in AdS$_3$, where conical deficits and BTZ black holes are related to empty AdS through a quotient. Effectively, this means we can solve a problem in empty flat space and map the result to the flat space cosmology through a non-single valued coordinate transformation. The diffeomorphism relating the FSC to Minkowski space reads
\be\label{eq:FSCMINK}
\begin{split}
r&=\sqrt{M(t^2-x^2)+R_c^2}\, , \\
\phi&=-{1\over{\sqrt{M}}} \log{{\sqrt{M}(t-x)}\over{r+R_c}}\, , \\
u&={1\over M} \left(   r-\sqrt{M} y-\sqrt{M} R_c \phi    \right) \, .
\end{split}
\ee
Using this map, we can relate the problem in the FSC to a problem in Minkowski space. A picture of the coordinate systems we are going to use in this section can be found in figure \ref{fig:setup}.   

We first consider the null lines falling from the boundary at the locations of the light primaries. These null lines have fixed values $u_i$ and $\phi_i$. Under the map \ref{eq:FSCMINK}, they correspond to the following lines in Minkowski space
\be\label{eq:SOLXI}
\begin{split}
x_{\gamma_i}&={{\lambda_i \sinh\left(\sqrt{M}\phi_i\right)-R_c \cosh\left(\sqrt{M}\phi_i\right)}\over{\sqrt{M}}}\, , \\
y_{\gamma_i}&={{\lambda_i -M u_i}\over{\sqrt{M}}}-R_c \phi_i\, , \\
t_{\gamma_i}&={{\lambda_i \cosh\left(\sqrt{M}\phi_i\right)-R_c \sinh\left(\sqrt{M}\phi_i\right)}\over{\sqrt{M}}}\, .
\end{split}
\ee
The horizon of the FSC is located at $r=R_c$ in the coordinates \ref{eq:ds}. This corresponds to the union of two planes in Minkowski space-time
\be\label{eq:FSCMINK}
{\cal H}=N_+ \cup N_-\, , \quad N_{\pm}:  t_{\cal H}=\pm x_{\cal{H}}\, .
\ee
The intersection of the planes $N_{\pm}$ is special, and we will denote it as
\be\label{eq:FSCMINK}
\gamma_{{\cal H}}=N_+ \cap N_- \, : \, t_{\cal H}=x_{\cal{H}}=0\, .
\ee
Our prescription now consists on introducing a geodesic network connecting the null lines ${\gamma_i}$ with the special horizon geodesic $\gamma_{\cal H}$. The simplest nework consists of three straight lines connected at a common bulk vertex point that we will denote $\mathbf{x}_v$. In this subsection, we consider spinless light operators, which should correspond to spinless bulk probes. This means that the on-shell action is the length of the particles' worldlines
\be\label{eq:Sspinless}
S=\xi_{\alpha}L(\mathbf{x}_{{\cal H}},\mathbf{x}_v)+\xi_{L} \left( L(\mathbf{x}_v,\mathbf{x}_{\gamma_1})  +L(\mathbf{x}_v,\mathbf{x}_{\gamma_2})   \right)\, ,
\ee
where $\xi_{\alpha}$ is the mass of the exchanged probe, and $\xi_L$ is the mass of the probes falling from the boundary. The lengths of the different geodesics are the standard length in Minkowski space
\be
L(x,x')=\sqrt{-(t-t')^2+(x-x')^2+(y-y')^2}\, .
\ee
The on-shell action \ref{eq:Sspinless} now has to be extremized with respect to the free parameters of the problem. These are the location of the common vertex point $\mathbf{x}_v$, the location of $\mathbf{x}_{\gamma_{\cal H}}$ along the horizon geodesic $\gamma_{\cal H}$, and the location of the points $\mathbf{x}_{\gamma_i}$ along the null geodesics $\gamma_i$.

The first step in this calculation is to extremize $S$ with respect to $y_{\cal H}$. This yields $y_{\cal H}=y_{v}$. We are left with
\be
S=\xi_{\alpha}\sqrt{-t_v^2+x_v^2}+\xi_{L} \left( L(\mathbf{x}_v,\mathbf{x}_{\gamma_1})  +L(\mathbf{x}_v,\mathbf{x}_{\gamma_2})   \right)\, .
\ee
Extremizing with respect to $\mathbf{x}_{\gamma_1}$ and $\mathbf{x}_{\gamma_2}$ yields a system of two equations that determine $x_v$ and $y_v$. The explicit solution reads
\be\label{eq:SOLXV}
\begin{split}
x_v&={{       \sqrt{M}(u_1-u_2) +R_c(\phi_2-\phi_1)+t_v \left(     \cosh{\sqrt{M}\phi_1} -\cosh{\sqrt{M}\phi_2}    \right)              }\over{     \sinh{\sqrt{M}\phi_2} - \sinh{\sqrt{M}\phi_1}            }}\, ,\\
y_v &={{    t_v \sinh {\sqrt{M}(\phi_1-\phi_2)}+R_c(\phi_2\sinh\sqrt{M}\phi_1  -\phi_1\sinh\sqrt{M}\phi_2  )   +\sqrt{M}\left(   u_2\sinh{\sqrt{M}\phi_1}  -u_1\sinh{\sqrt{M}\phi_2}     \right)         }\over{     \sinh{\sqrt{M}\phi_2} - \sinh{\sqrt{M}\phi_1}            }}\, .
\end{split}
\ee
Replacing these expressions back in the on-shell action yields
\be
\begin{split}
S=&\xi_L (u_1-u_2)\coth{\left[{1\over 2}\sqrt{M}(\phi_1-\phi_2)\right]} +R_c\left[ -{2\over{\sqrt{M}}}  +(\phi_1-\phi_2)\coth\left( {{\sqrt{M}}\over 2} (\phi_1-\phi_2)\right) \right]\\
&+ \xi_{\alpha}\sqrt{\left(    {{    \sqrt{M}(u_1-u_2) +R_c(\phi_2-\phi_1)+t_v \left(     \cosh{\sqrt{M}\phi_1} -\cosh{\sqrt{M}\phi_2}    \right)       }\over{    \sinh{\sqrt{M}\phi_2} - \sinh{\sqrt{M}\phi_1}    }}       \right)^2-t_v^2}\, .
\end{split}
\ee
Extremizing with respect to $t_v$ results in
\be\label{eq:SOLXV2}
t_v={{\sqrt{M}(u_1-u_2)+R_c(\phi_1-\phi_2)}\over{2}}{{e^{\sqrt{M}(\phi_1+\phi_2)}-1}\over{e^{\sqrt{M}\phi_1}-e^{\sqrt{M}\phi_2}}}  \, ,
\ee
which leads to the extremal value of $S$
\be\label{eq:SspinlessAUX}
\begin{split}
S=&\xi_L \left( (u_1-u_2)\coth{\left[{1\over 2}\sqrt{M}(\phi_1-\phi_2)\right]} +R_c\left[ -{2\over{\sqrt{M}}}  +(\phi_1-\phi_2)\coth\left( {{\sqrt{M}}\over 2} (\phi_1-\phi_2)\right) \right]\right)\\
&+ \xi_{\alpha}{{e^{\sqrt{M}(\phi_1+\phi_2)}}\over{e^{\sqrt{M}\phi_1}-e^{\sqrt{M}\phi_2}}}\left( \sqrt{M} (u_1-u_2)+R_c(\phi_1-\phi_2)  \right)\, .
\end{split}
\ee
This result is written in the cylinder coordinates of \ref{eq:ds}. The computations done in section \ref{sec:BMSFT} use coordinates in the plane. In order to compare both results, we perfom the following map
\be\label{eq:PlaneCyl}
U=e^{i \phi}\, , \quad \text{and} \quad V=i u e^{i \phi}\, .
\ee
The map transforms the primaries non-trivially, as in equation \ref{eq:PrimaryTransformation}. Taking this into account, and making the replacements $U_1=1$, $V_1=0$, $U_2=U$, and $V_2=V$ yields the result
\be
S_{\text{extr}}=   V\left(  {{\tilde{\beta} U^{{{\tilde{\beta}}\over 2}}}\over{U(U^{\tilde{\beta}}-1)}}\xi_{\alpha} - {{U^{\tilde{\beta}}(\tilde{\beta}+1)+\tilde{\beta}-1}\over{U(U^{\tilde{\beta}}-1)}}\xi_L  \right) +\tilde{\delta}_H\log U \left(     {{2U^{{\tilde{\beta}}\over 2}}\over{\tilde{\beta} (U^{\tilde{\beta}}-1)}}    \xi_{\alpha}            +  {{2(U^{\tilde{\beta}}+1)}\over{\tilde{\beta}(1-U^{\tilde{\beta}})}} \xi_L                    \right) \, .
\ee
Here we have defined $\tilde{\beta}=\sqrt{-M}$, and $\tilde{\delta}_H=J/2\sqrt{M}$. The result found in this section matches the exponential part of the BMSFT result found in equation \ref{eq:ResultBMSFT} under the following identification
\be
\tilde{\beta}=\beta\, , \quad \text{and}\quad \tilde{\delta}_H=\delta_H\, .
\ee
In terms of the quantum numbers of the heavy operators, this implies
\be
M={{24\xi_H}\over{c_M}}-1\, , \quad  \text{and}\quad {J\over{2\sqrt{M}}}={{6\Delta_H}\over{c_M}}\, .
\ee
We conclude that a pair of Heavy operators in the null plane corresponds to a FSC only if the quantum number $\xi_H$ is greater than $c_M/24$. If the quantum number is below this bound, the mass of the FSC becomes negative. In this case, the background solution would correspond to a conical deficit or surplus \cite{Barnich:2012aw}.  

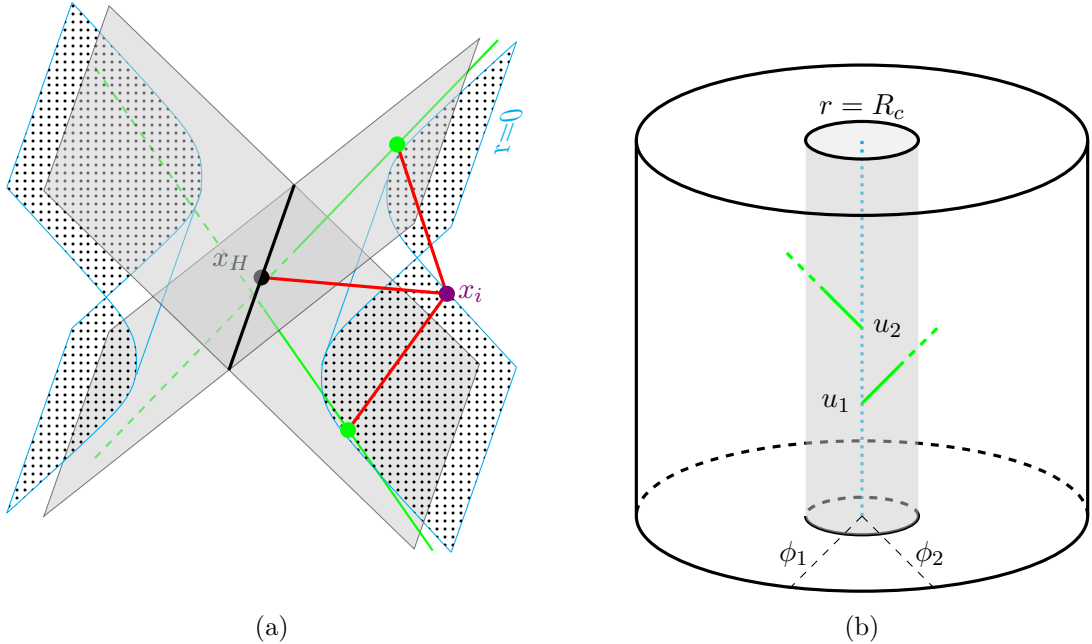
\begin{figure}[t!]
\centering
\begin{subfigure}[t]{0.48\textwidth}
        \centering

\tdplotsetmaincoords{60}{100}
\begin{tikzpicture}[scale=5,tdplot_main_coords]

    \def\x{1}

  \filldraw[
        draw=black,%
        fill=black!20,
         opacity=0.5,%
    ]          (0,1,0)
            -- (1,1,0)
            -- (1,1/2,1/2)
            -- (0,1/2,1/2)
            -- cycle;

\draw [green, thick] (0,1+0.05,1)--(1/4,1/2+0.05,1/2-0.05  );
\draw [green, thick,dashed] ( 1/2,0.05,-0.1      )--(1/4,1/2+0.05,1/2-0.05  );

\draw [green,thick] (1,1+0.05,0)--(  3/4,1/2+0.05,1/2+0.05 );
\draw [green,thick,dashed] (1-2/4,0.05,1.1)--(  3/4,1/2+0.05,1/2+0.05 );

\draw [cyan,name path=Lpath] plot [smooth, tension=0.5] coordinates { (1,1+0.1,0) (1,1/2+0.3,1/2-0.15)  (1,1/2+0.25,1/2) (1,1/2+0.3,1/2+0.15) (1,1+0.1,1)};
\draw [cyan] (1,1+0.1,0)--(0,1+0.1,0);
\draw [cyan,name path=Rpath] plot [smooth, tension=0.5] coordinates { (0,1+0.1,0) (0,1/2+0.3,1/2-0.15)  (0,1/2+0.25,1/2) (0,1/2+0.3,1/2+0.15) (0,1+0.1,1)};
\draw [cyan,name path=Upath] (1,1+0.1,1)--(0,1+0.1,1)  node[midway,sloped,below,xslant=-0.5] {r=0};
\tikzfillbetween[
    of=Lpath and Rpath,split
  ] {pattern=dots};

\draw [cyan,name path=Mpath] (0,1/2+0.25,1/2+0.04)--(1,1/2+0.25,1/2+0.04);

    \filldraw[
        draw=black,%
        fill=black!20,
        opacity=0.5,%
    ]          (0,0,0)
            -- (1,0,0)
            -- (1,1,1)
            -- (0,1,1)
            -- cycle;

\draw [cyan,name path=Lpath2] plot [smooth, tension=0.5] coordinates { (1,0-0.1,0) (1,1/2-0.3,1/2-0.15)  (1,1/2-0.25,1/2) (1,1/2-0.3,1/2+0.15) (1,0-0.1,1)};
\draw [cyan] (1,0-0.1,0)--(0,0-0.1,0);
\draw [cyan,name path=Rpath2] plot [smooth, tension=0.5] coordinates { (0,0-0.1,0) (0,1/2-0.3,1/2-0.15)  (0,1/2-0.25,1/2) (0,1/2-0.3,1/2+0.15) (0,0-0.1,1)};
\draw [cyan,name path=Upath] (1,0-0.1,1)--(0,0-0.1,1);

\tikzfillbetween[
    of=Lpath2 and Rpath2,split
  ] {pattern=dots};
\draw [cyan,name path=Mpath] (0,1/2-0.25,1/2-0.04)--(1,1/2-0.25,1/2-0.04);

\draw [red, very thick,-dotb-=1] (1/2,1,1/2)--(  1/2,1/2,1/2) node[pos=1,above=6,left]{\textcolor{black}{$x_H$}} node[pos=0,right]{\textcolor{violet}{$x_i$}};
\draw [red, very thick,-dotg-=1] (1/2,1,1/2)--(  7/8, 3/4+0.05,1/4+0.025 );
\draw [red, very thick,-dotp-=0,-dotg-=1] (1/2,1,1/2)--(  1/8,3/4+0.05,3/4-0.025 );

  \filldraw[
        draw=black,%
        fill=black!20,
         opacity=0.5,%
    ]          (0,0,1)
            -- (1,0,1)
            -- (1,1/2,1/2)
            -- (0,1/2,1/2)
            -- cycle;

\draw (1,0,0) node[below] {\textcolor{white}{$N_2$}};
\draw (1,1,0) node[below] {\textcolor{white}{$N_1$}};

  \draw[
       black,very thick%
    ]          (0,1/2,1/2)
            -- (1,1/2,1/2);

\end{tikzpicture}

        \caption{ }
    \end{subfigure}%
\quad
\begin{subfigure}[t]{0.48\textwidth}
        \centering

\begin{tikzpicture}

 \fill[color=black!20, opacity=0.25](2,7) ellipse (0.75 and 0.25);

\draw[white,very thick,name path=TOP](1.25,7) arc (180:360:0.75 and 0.25);

\draw[cyan,very thick,dotted] (2,2) -- (2,7) node [pos=1,above=2]{{$ $}} node [pos=0,below=2]{{$ $}};




\draw[black,very thick,dashed](2.75,2) arc (0:180:0.75 and 0.25);
\draw[black,very thick,name path=BOT](1.25,2) arc (180:360:0.75 and 0.25);

\draw[black,very thick,dashed](5,2) arc (0:180:3 and 1);
\draw[black,very thick](-1,2) arc (180:360:3 and 1);




 \fill[color=black!20, opacity=0.5]
(1.25,7)--(1.25,2)
-- (1.25,2) arc (180:360:0.75 and 0.25)
-- (2.75,7) 
-- (2.75,7) arc (360:180:0.75 and 0.25) --cycle;

\draw[green,very thick] (2,4.5) -- (1.5,5) node[pos=0,right]{\textcolor{black}{$u_2$}};
\draw[green,very thick,dashed] (1,5.5) -- (1.5,5);

\draw[green,very thick] (2,3.5) -- (2.5,4) node[pos=0,left]{\textcolor{black}{$u_1$}};
\draw[green,very thick,dashed] (3,4.5) -- (2.5,4);

\draw[black,very thick] (-1,2) -- (-1,7);
\draw[black,very thick] (5,2) -- (5,7);

\draw[black,very thick](2,7) ellipse (0.75 and 0.25);

\draw[black,very thick](2,7) ellipse (3 and 1)  node [above=4] {$r=R_c$};

\draw[black,dashed](2,2) --(3,1) node [pos=0.5,right=2] {$\phi_2$};
\draw[black,dashed](2,2) --(1,1) node [pos=0.5,left=2] {$\phi_1$};

\end{tikzpicture}


        \caption{ }
    \end{subfigure}
    \caption{a) Holographic BMS$_3$ block in the Minkowski coordinates \ref{eq:FSCMINK}. The gray planes correspond to the horizon of the FSC. The singularity at $r=0$ in ADM coordinates corresponds to the surface $x^2-t^2=R_c^2/M$. The purple dot represents the vertex point $x_i$, whose location after extremizing the on-shell action is given explicitly in \ref{eq:SOLXI}. The common vertex point lies past the singularity.   b) FSC in ADM coordinates. The gray cylinder represents the horizon at $r=R_c$. The origin at $r=0$ is represented by a dashed blue line. The green null lines fall from the null boundary at the points $u_i,\phi_i$. }\label{fig:geompic}
\end{figure}

The result presented here matches the BMSFT result found in equation \ref{eq:FM}. It is worth mentioning that the geometric configuration given by the expressions for $\mathbf{x}_v$ in \ref{eq:SOLXV} and \ref{eq:SOLXV2} maps to a complex value of the $r,u,\phi$ coordinates under the inverse of the map \ref{eq:FSCMINK}. As shown in figure \ref{fig:geompic}, the vertex point lies in the region of Minkowski space not covered by the original geometry. Namely, the vertex lies in the region behind the coordinate singularity at $r=0$. This implies that the map \ref{eq:FSCMINK} results in a complex value of  $r,u,\phi$ for the point $\mathbf{x}_v$. This is a commonplace feature of the stationary phase approximation of correlators in Lorentzian space-times. The calculations in this section are expected to correspond to the stationary phase approximation of a path integral involving fields propagating in a background geometry. The approximation arises as one takes the heavy limit, where the fields become heavy enough to classicalize. The fact that the saddle is complex just indicates that a single path does not dominate the integral, but a family of them. The complex saddle simply captures the contributions from such family of dominant paths. A simple example where this phenomenon happens in the context of AdS/CFT  is the calculation of the holographic lorentzian two-point function of primary operators separated by a time-like interval. In this case the relevant computation consists on finding the length of the geodesic connecting the locations of the primary operators at the boundary. However, no such geodesic exists if the operators are time-like separated. The approximation in this case results on a complex geodesic whose length correctly computes the correlator, leading to the same kind of feature found in this section. For another more interesting example see \cite{Anous:2017tza}.

In this section we have studied massive spinless particles that propagate in a flat space cosmology. The on-shell action associated to such particles is precisely the length of the particle's worldline. We have thus extremized  the length of a geodesic network. In order to obtain the full answer for the block, we also have to study probe particles with spin.  We turn now to the study of massive spinning probes in a flat space cosmology. We expect the final result to match the full answer in equation \ref{eq:ResultBMSFT}.

\subsection{Spinning holographic BMS$_3$ blocks }
In this section we perform computations of holographic blocks involving operators with non-trivial $L_0$ quantum numbers. In previous work we concluded that the particles dual to such operators had non-trivial spin. Poincar\'e blocks where computed by considering the on-shell action of massive particles with a Lorentz frame attached. This construction was introduced in the AdS/CFT literature in \cite{Castro:2014tta}. The computations in this subsection will generalize the ones in the previous one by considering spinning light probes. 

The introduction of a Lorentz frame attached to each probe does not modify the $\xi$ dependent part of the calculation. We will thus consider the same set-up as last section, with the addition of the following extra terms to the on-shell action
\be
S^{\text{spin}}=\Delta_L \left[  \log\left( -2\dot{\gamma}_1 \cdot n_v \right)    +\log\left( -2\dot{\gamma}_2 \cdot n_v \right)        \right] +\Delta_{\alpha} \cosh\left(   -n_v \cdot n_{\cal H}  \right)\, .
\ee
Here $\Delta_L$ and $\Delta_{\alpha}$ are the $L_0$ quantum numbers of the external and exchanged operators respectively. $n_v$ is a time-like vector located at the common vertex point $x_v$. The vector $n_{\cal H}$ is also time-like and it is located at the point $x_{\cal H}$ perpendicularly to the special horizon geodesic $\gamma_{\cal H}$. The vectors $\dot{\gamma}_i$ are null and point along the null geodesics $\gamma_i$. For more details concerning this set-up we point the reader to the work \cite{Hijano:2017eii}. 

The expression for the null vectors $\dot{\gamma}_i$ in Minkowski space coordinates is
\be
\dot{\gamma}_i={{\cosh{\sqrt{M}\phi_i}}\over{\sqrt{M}}}\partial_t   +{{\sinh{\sqrt{M}\phi_i}}\over{\sqrt{M}}}\partial_x +{1\over{\sqrt{M}}}\partial_y\, .
\ee
A particularly useful parametrization of the time-like vectors is
\be
\begin{split}
n_v&={{z_v^2+u_v^2+1}\over{2u_v}}\partial_t+{{z_v^2+u_v^2-1}\over{2u_v}}\partial_x +{{z_v}\over{u_v}}\partial_y\, ,\\
n_{\cal H}&={{z_{\cal H}^2+u_{\cal H}^2+1}\over{2u_{\cal H}}}\partial_t+{{z_{\cal H}^2+u_{\cal H}^2-1}\over{2u_{\cal H}}}\partial_x +{{z_{\cal H}}\over{u_{\cal H}}}\partial_y\, .
\end{split}
\ee
The parameters $u$ and $z$ can be associated to coordinates in AdS$_2$. However, we will not explore the relation between this computation and AdS$_2$ gravity calculations in this note. In order for the vector $n_{\cal H}$ to be perpendicular to the horizon, we must set $x_{\cal H}=0$. Writing the spinning action explicitly and extremizing with respect to $u_{\cal H}$ yields the solution
\be
u_{\cal H}=\sqrt{u_v^2+x_v^2}\, .
\ee
Replacing this expression back into the action results in
\be
\begin{split}
S^{\text{spin}}&=\Delta_{\alpha} \log\left(    {{\sqrt{u_v^2+x_v^2}}\over{u_v}}+{{x_v}\over{u_v}} \right)\\
&+\Delta_L \left(   \log\left[     {{u_v^2+\left(e^{\sqrt{M}\phi_1}-x_v\right)^2  }\over{e^{\sqrt{M}\phi_1} M u_v}}    \right]   +\log\left[     {{ u_v^2+\left(e^{\sqrt{M}\phi_2}-x_v\right)^2   }\over{e^{\sqrt{M}\phi_2} M u_v}}   \right]    \right)\, .
\end{split}
\ee
We find it useful to replace variables as $u_v=2 K x_v/(K^2-1)$. Extremizing with respect to $x_v$ now yields
\be
x_v=e^{{{\sqrt{M}}\over 2}\left(  \phi_1+\phi_2  \right)}{{K^2-1}\over{K^2+1}}\, .
\ee
Replacing this solution in the on-shell action and exremizing with respect to the remaining variable $K$ yields
\be
K=\sqrt{{{2\Delta_L-\Delta_{\alpha}}\over{2\Delta_L+\Delta_{\alpha}}}} {{  e^{  {{\sqrt{M}}\over{2}}\phi_1  } +e^{  {{\sqrt{M}}\over{2}}\phi_2  }    }\over{  e^{  {{\sqrt{M}}\over{2}}\phi_1  } -e^{  {{\sqrt{M}}\over{2}}\phi_2  }    }}\, .
\ee
The final answer for the extremized spinning action after changing to plane coordinates as in \ref{eq:PlaneCyl}, reads
\be
S^{\text{spin}}_{\text{extr}}= \Delta_{\alpha} \log \left( {{1-U^{{\tilde{\beta}}\over{2}}}\over{1+U^{{\tilde{\beta}}\over{2}}}}  \right) +\Delta_L \log \left( {{   U^{\tilde{\beta}-1}    }\over{  (1-U^{\tilde{\beta}})^2    }}   \right)\, .
\ee
Where $\tilde{\beta}$ is defined as in the last subsection. This result, added to the one obtained in the last subsection, completes our holographic calculation of a BMS$_3$ block. The result successfully matches the full BMSFT result found previously in \ref{eq:ResultBMSFT}.

The computation presented in this section can be understood as evidence that the extrapolate dictionary proposed in \cite{Hijano:2017eii} and explained in formula \ref{eq:Extrapolate} applies to geometries beyond empty Minkowski space. We have also shown that the field theory states resulting from the insertion of heavy primary BMS operators in the null plane are dual to flat space cosmologies in the bulk. This is a statement that seems to hold at the level of the correlators of light operators. It is tempting to interpret this statements as some version of the ETH in flat holography.

\section{Discussion}\label{sec:discussion}

We have derived a closed form expression for semi-classical BMS$_3$ blocks, extending the results in \cite{Bagchi:2017cpu}. The field theory techniques revolve around the monodromy method, which has been used successfully to compute Virasoro blocks in the AdS/CFT literature. The application of the monodromy method to theories with BMS symmetry requires the introduction of reducible but indecomposable representations we have dubbed as ``BMS$_3$ multiplets". These multiplets in principle do not need to be part of the spectrum of the theory, but it is worth mentioning that level two descendants of the identity operator constitute one of these multiplets. We believe it would be of interest to explore these representations in more detail in future work. 

Holographically, we have achieved a clean flat space bulk interpretation of the semi-classical blocks, generalizing the observations in \cite{Hijano:2015rla,Hijano:2015zsa} to flat space. This is evidence supporting the extrapolate dictionary conjectured  in \cite{Hijano:2017eii}. The computations we have performed rely on the probe limit $1\ll \Delta_{L},\xi_L \ll c_M$. It would be interesting to test the dictionary away from the probe limit. This yields two possible directions of future work. On one hand, one could study heavier probes. In principle, the bulk picture can be analyzed to higher orders in the light operator dimensions. This would correspond to solving Einstein's equations order by order in the Newton's constant. In the field theory, one would need to solve the monodromy method to higher order in $\xi_L/c_M$ and $\Delta_L/c_M$. On the other hand, it would be interesting to understand the calculation for light operator dimensions $\Delta_L,\xi_L\sim {\cal O}(1)$.

The fact that the blocks can be obtained in the bulk by studying probes propagating in a thermal background can be interpreted as the statement of ETH. In the field theory, the insertion of heavy operators yields a pure, highly energetic  microstate. However, in this note we have shown that the blocks involving low energy observables in such microstate can be computed by studying bulk probes propagating in a thermal background geometry in the semi-classical limit. 

The analysis presented in this work does not require any knowledge of conformal field theory results, and it is based exclusively on the structure of the BMS algebra. However, the final expression for the blocks can also be obtained from the non-relativistic limit of Virasoro conformal blocks. The details concerning the non-relativistic limit can be found in appendix \ref{app:limit}.

\acknowledgments
I thank Tarek Anous, Charles Rabideau, and Francis Duplessis for very useful discussions. I wish to acknowledge support
from the Simons Foundation through the It-from-Qubit collaboration.

\appendix

\section{Four-point function - Monodromy of ${\cal M}$ }\label{app:Mmon}
We start from the differential equation written in \ref{eq:diffO13} which we re-write here
\be
g_i^{(1)\prime\prime}+{6\over {c_M}}{\cal M}^{(0)}(u_3;u)g_i^{(1)}=-{6\over {c_M}}{\cal M}^{(1)}(u_3;u)g_i^{(0)}\, .
\ee
with
\be
\begin{split}
g_i&=g_i^{(0)}+\epsilon_{\alpha} g_i^{(1)}\, , \\
{6\over {c_M}}{\cal M}(u_3;u)&={1\over{u^2}}{{1-\beta^2}\over 4} -{{u_3}\over{u(u-u_3)^2}}\epsilon_{\alpha}={6\over {c_M}}{\cal M}^{(0)}(u_3;u)+\epsilon_{\alpha}{6\over {c_M}}{\cal M}^{(1)}(u_3;u)  \, .
\end{split}
\ee
The solutions to this differential equation at zeroth order in $\epsilon_{\alpha}$  are simply
\be
g_1^{(0)}(u)=u^{{1+\beta}\over{2}}\, , \quad \text{and}\quad g_2^{(0)}(u)=u^{{1-\beta}\over{2}}\, .
\ee
The Wronskian associated to these solutions reads
\be
W=\left|\begin{matrix}
   g_1^{(0)}(u) & g_1^{(0)\prime}(u) \\
 g_2^{(0)}(u) & g_2^{(0)\prime}(u) 
\end{matrix}\right|= -\beta\, .
\ee
It will also be necessary to define the modified Wronskians, whose explicit form is
\be
W_{i,1}=\left|\begin{matrix}
   0 & -{6\over {c_M}}{\cal M}^{(1)}(u_3;u)g_i^{(0)} \\
 g_2^{(0)}(u) & g_2^{(0)\prime}(u) 
\end{matrix}\right| \, , \quad \text{and}\quad W_{i,2}=\left|\begin{matrix}
    g_1^{(0)}(u) & g_1^{(0)\prime}(u) \\
 0 & -{6\over {c_M}}{\cal M}^{(1)}(u_3;u)g_i^{(0)} 
\end{matrix}\right|\, .
\ee
We are now ready to write the functions $f_{i,j}$ appearing in the solution to the inhomogeneous equation
\be
g_i^{(1)}(u)=f_{i,1}(u)g_1^{(0)}(u)+f_{i,2}(u)g_2^{(0)}(u)\, ,
\ee
with
\be
f^{\prime}_{i,j}(u)={{W_{i,j}(u)}\over{W(u)}}\, .
\ee
The explicit expressions read
\be\label{eq:AppAfij}
\begin{split}
f^{\prime}_{1,1}(u)&={{U(U-1)}\over{(u-1)(u-U)\beta}}c_U   +{{U^2(u-2)+u+2u U(u-2)}\over{(u-U)^2(u-1)^2\beta}}\epsilon_{L}\, , \\
f^{\prime}_{1,2}(u)&=-{{u^{\beta}U(U-1)}\over{(u-1)(u-U)\beta}}c_U   + {{u^{\beta+1}\left(   {{u-2}\over{u(u-1)^2}}    -  {{1}\over{(u-U)^2}}       \right)}\over{\beta}}  \epsilon_{L} \, , \\
f^{\prime}_{2,1}(u)&={{u^{-\beta}U(U-1)}\over{(u-1)(u-U)\beta}}c_U   + {{U^2(u-2)+u+2u U(u-2)}\over{u^{\beta}(u-U)^2(u-1)^2\beta}} \epsilon_{L} \, , \\
f^{\prime}_{2,2}(u)&= -{{U(U-1)}\over{(u-1)(u-U)\beta}}c_U    +   {{   U^2(u-2)-u-2uU(u-2)     }\over{(u-U)^2(u-1)^2\beta}}   \epsilon_L             \, .
\end{split}
\ee
We are interested only on how the functions $f_{i,j}$ behave as we move our coordinate $u,v$ around the positions of the light operators at $(u_3,v_3)=(1,0)$ and $(u_4,v_4)=(U,V)$. It is obvious from expressions \ref{eq:AppAfij} that the functions $f_{1,1}$ and $f_{2,2}$ remain the same, due to the lack of non-trivial branch structure. The functions $f_{1,2}$ and $f_{2,1}$ are more complicated due to the presence of non integer powers of $u$. An efficient way of computing the way these functions change as we move around some closed cycle is to compute the residues of the corresponding contour integral. The result yields
\be
\begin{split}
\int_{{\cal C}(x_3,x_4)} f'_{1,2} &= -2\pi i \left(  {{U(U^{\beta}-1)}\over{\beta}} c_U  +{{U^{\beta}(\beta+1)+\beta-1}\over{\beta}}\epsilon_L      \right)\, , \\
\int_{{\cal C}(x_3,x_4)} f'_{2,1} &= 2\pi i \left( {{U(U^{-\beta}-1)}\over{\beta}}c_U  -{{U^{\beta}(\beta+1)+\beta-1}\over{\beta U^{\beta}}}\epsilon_L \right) \, .
\end{split}
\ee
These results explicitly show how the solutions $g_i$ of the differential equation \ref{eq:diffg} transform when we move the insertion of the current operator around the locations of the light primaries. Namely
\be
\begin{split}
g_1(u)&\rightarrow g_1(u) +\left( \int_{{\cal C}(x_3,x_4)} f'_{1,2}\right) g_2(u)\, ,\\
g_2(u)&\rightarrow \left( \int_{{\cal C}(x_3,x_4)} f'_{2,1}\right) g_1(u)+g_2(u) \, .
\end{split}
\ee
In the language of the transformation of the coordinate transformation $P(u)$, the result for the contour integrals performed here explicitly yield expressions for the parameters $A,B,C,D$ in
\be
P(u)={{g_1(u)}\over{g_2(u)}} \xrightarrow{C_{(x_3,x_4)}} {{A g_1(u) +B g_2(u)}\over{C g_1(u)+D g_2(u)}}={{A P(u)+B}\over{C P(u) +D}}\, .
\ee
We can thus easily combine our results into the invariant objects $I_1$ and $I_2$, which are defined as
\be
\begin{split}
I_1=\text{tr}M\, , \quad \text{and}\quad I_2=\text{tr}M^2\, ,
\end{split}
\ee
where we have defined the monodromy matrix
\be
M=\left(\begin{matrix}
   A &B\\
  C    & D 
\end{matrix}\right)\, .
\ee
The result is
\be
\begin{split}
I_1&=2\, , \\
\sqrt{{{I_1-I_2}\over 2}}& =2\pi {{U^{-{{\beta}\over 2}}}\over{\beta}} \left[ c_4 U(U^{\beta}-1)+\epsilon_{L}\left( U^{\beta}(\beta+1)+\beta-1    \right)   \right]\, .
\end{split}
\ee
These expressions were used in section \ref{sec:Mmon4} to compute part of the BMS$_3$ block. 


\section{Three-point function - Monodromy of $\cal L$}\label{app:MONODROMYL3}
In this appendix we study the monodromy of the differential equation \ref{eq:diffeqX}.  To zeroth order in $\epsilon_{\alpha}$ and $\delta_{\alpha}$ the three solutions read
\be\label{eq:X0}
\begin{split}
X_1^{(0)}&=2{{\delta_H}\over{\beta^2}}u \log u +{1\over{P'}}\, ,\\
X_2^{(0)}&=2{{\delta_H}\over{\beta^2}}u \log u +{P\over{P'}}\, , \\
X_3^{(0)}&=2{{\delta_H}\over{\beta^2}}u \log u +{{P^2}\over{P'}}\, .
\end{split}
\ee
In order to solve the problem to linear oder in $\epsilon_{\alpha}$ and $\delta_{\alpha}$, we expand $X(u)$ around the solutions \ref{eq:X0} and write down the currents as ${\cal L}^{(0)}+{\cal L}^{(1)}$ and ${\cal M}^{(0)}+{\cal M}^{(1)}$. The resulting differential equation reads
\be\label{eq:X1}
X_i^{(1)\prime\prime\prime}-4X_i^{(1)\prime}{6\over{c_M}}{\cal M}^{(0)}-2X_i^{(1)}{6\over{c_M}}{\cal M}^{(0)\prime}=-2{6\over{c_M}}\left(  {\cal L}^{(1)}-X_i^{(0)}{\cal M}^{(1)\prime}-2X_i^{(0)\prime}{\cal M}^{(1)}  \right)\, .
\ee
with
\be
\begin{split}
{6\over{c_M}}{\cal L}^{(0)}&={{\delta_H}\over{u^2}} \, , \quad {6\over{c_M}}{\cal L}^{(1)}={{u_3}\over{(u-u_3)^2}}\delta_{\alpha} -{{u v_3(u+u_3)}\over{u(u-u_3)^3}}\epsilon_{\alpha}\, , \\
{6\over{c_M}}{\cal M}^{(0)}&={1\over{u^2}}{{1-\beta^2}\over 4}\, , \quad \text{and}\quad {6\over{c_M}}{\cal M}^{(1)}={{u_3}\over{u(u-u_3)^2}}\epsilon_{\alpha} \, .
\end{split}
\ee
The solutions can be written as
\be
X_i^{(1)}=f_{i,1}X_1^{(0)}+f_{i,2}X_2^{(0)}+f_{i,3}X_3^{(0)}\, ,
\ee
where again, the functions $f_{i,j}$ read
\be\label{eq:fprimeL}
f^{\prime}_{i,j}(u)={{W_{i,j}(u)}\over{W(u)}}\, .
\ee
Here $W(u)$ is the Wronskian built out of the family of solutions $X_i^{(0)}$
\be\label{eq:Wapp}
W(u)=\left|\begin{matrix}
X_1^{(0)} &  X_1^{(0)\prime}& X_1^{(0)\prime\prime} \\
X_2^{(0)} &  X_2^{(0)\prime}& X_2^{(0)\prime\prime} \\
X_3^{(0)} &  X_3^{(0)\prime}& X_3^{(0)\prime\prime} 
\end{matrix}\right|\, .
\ee
The objects $W_{i,j}(u)$  are the Wronskian, with the $j^{\text{th}}$ row modified by the inhomogeneous piece of   the differential equation \ref{eq:X1}.
\begin{equation*}
W_{i,1}(u)=\left|\begin{matrix}
0 & 0& Z_i\\
X_2^{(0)} &  X_2^{(0)\prime}& X_2^{(0)\prime\prime} \\
X_3^{(0)} &  X_3^{(0)\prime}& X_3^{(0)\prime\prime} 
\end{matrix}\right|\, , \quad W_{i,2}(u)=\left|\begin{matrix}
X_1^{(0)} &  X_1^{(0)\prime}& X_1^{(0)\prime\prime} \\
0&  0 & Z_i \\
X_3^{(0)} &  X_3^{(0)\prime}& X_3^{(0)\prime\prime} 
\end{matrix}\right|\, , \quad  W_{i,3}(u)=\left|\begin{matrix}
X_1^{(0)} &  X_1^{(0)\prime}& X_1^{(0)\prime\prime} \\
X_2^{(0)} &  X_2^{(0)\prime}& X_2^{(0)\prime\prime} \\
0& 0& Z_i
\end{matrix}\right|\, , 
\end{equation*}
with
\be\label{eq:ziapp}
Z_i =-2{6\over{c_M}}\left(  {\cal L}^{(1)}-X_i^{(0)}{\cal M}^{(1)\prime}-2X_i^{(0)\prime}{\cal M}^{(1)}  \right)\, .
\ee
In principle, in order to write down the full solution we would need to integrate equations \ref{eq:fprimeL}. However, we are only concerned with the monodromy of the solution when we encircle the operator $L_{\alpha}(x_3)$. This means we are only interested in the residues of the integrand shown in \ref{eq:fprimeL} at the point $u=u_3$. Once the residues are computed, we know how the solution for $G$ changes as we move around  $L_{\alpha}(x_3)$. Namely,
\be
\begin{split}
{{G}\over{P'}}&\xrightarrow{C_{\alpha}} {{G}\over{P'}}+ x_1 {1\over{P'}}+ x_2 {P\over{P'}}+ x_3 {{P^2}\over{P'}}\, , \\
P&\xrightarrow{C_{\alpha}} {{A P +B}\over{C P+D}}\, ,
\end{split}
\ee
where $A,B,C,$ and $D$ are the elements of the monodromy of ${\cal M}$ we computed in previous sections, and $x_i$ are the components of the ${\cal L}$ monodromy that can be computed through the residues of \ref{eq:fprimeL}. 

As when studying the monodromy of ${\cal M}$, the objects $x_i$ are not gauge invariant information. We need to understand what combinations of these objects are invariant under global BMS transformations. Before encircling the operator $L_{\alpha}$, say we perform a global transformation $g$
\be\label{eq:Lg}
\begin{split}
{{G}\over{P'}}&\xrightarrow{g} {{G}\over{P'}}+ y_1 {1\over{P'}}+ y_2 {P\over{P'}}+ y_3 {{P^2}\over{P'}}\, , \\
P&\xrightarrow{g} {{x P + y}\over{z P + t}}\, .
\end{split}
\ee
Once the coordinate system has been changed by applying $g$, we circle around the light operator and pick up a monodromy
\be
\begin{split}
{{G}\over{P'}}&\xrightarrow{C_{\alpha}g} {{G}\over{P'}}+ x_1 {1\over{P'}}+ x_2 {P\over{P'}}+ x_3 {{P^2}\over{P'}}+ y_1 {1\over{\left(  {{A P +B}\over{C P+D}} \right)'}}+ y_2 {\left(  {{A P +B}\over{C P+D}} \right)\over{\left(  {{A P +B}\over{C P+D}} \right)'}}+ y_3 {{\left(  {{A P +B}\over{C P+D}} \right)^2}\over{\left(  {{A P +B}\over{C P+D}} \right)'}}\, , \\
P&\xrightarrow{C_{\alpha}g} {{x\left(  {{A P +B}\over{C P+D}} \right) + y}\over{z \left(  {{A P +B}\over{C P+D}} \right) + t}}\, .
\end{split}
\ee
We then undo the global transformation by applying $g^{-1}$. The parameters defining $g^{-1}$ can be derived by applying a generic global transformation to \ref{eq:Lg} and demanding that both $P$ and $G$ return to themselves. The full transformation reads
\be\label{eq:LgiCg}
\begin{split}
{{G}\over{P'}}&\xrightarrow{g^{-1}C_{\alpha}g} {{G}\over{P'}}+ x'_1 {1\over{P'}}+ x'_2 {P\over{P'}}+ x'_3 {{P^2}\over{P'}}\, , \\
P&\xrightarrow{g^{-1}C_{\alpha}g} {{A' P + B'}\over{C' P + D'}}\, .
\end{split}
\ee
The primed parameters are functions of the parameters defining $g$, and the parameters defining the monodromy around the light primary. The non-trivial dependence on $g$ reveals the fact that the parameters $x_i$ and $A,B,C,D$ are not gauge invariant objects. However, one can find combinations which do not depend on $g$. Namely,
\be\label{eq:I3def}
I_3=-2 C x_1 +(D-A)x_2 +2B x_3\, .
\ee
This object does not depend on $g$, in the sense that
\be
I_3=-2 C x_1 +(D-A)x_2 +2B x_3=-2 C' x'_1 +(D'-A')x'_2 +2B' x'_3\, .
\ee
In the case of the monodromy of the current along the primary $L_{\alpha}(x_3)$, this invariant quantity reads
\be\label{eq:I3app}
\sqrt{I_3}=4\pi \sqrt{{{\delta_{\alpha}\epsilon_{\alpha}}\over{\beta}}}\, .
\ee
This concludes the computation of the gauge invariant part of the monodromy associated to the current ${\cal L}$ in the case of the three-point correlator.


\section{Four-point function - Monodromy of $\cal L$} \label{app:MONODROMYL4}
In this appendix we study the monodromy problem posed by the differential equation \ref{eq:diffeqX}. This time the expectation value of the current ${\cal L}$ is given in equation \ref{eq:Ltilde4}, while the expectation value of ${\cal M}$ is given in \ref{eq:M4}. To zeroth order in the light operators' quantum numbers, the solutions to the differential equation are the same as the one we found in appendix \ref{app:MONODROMYL3}. The space of solutions to zeroth order can thus be found in equation \ref{eq:X0}. The solutions to linear oder can be obtained as in appendix \ref{app:MONODROMYL4}, but this time using
\be
\begin{split}
{6\over{c_M}}{\cal L}^{(0)}&={{\delta_H}\over{u^2}} \, , \\
{6\over{c_M}}{\cal L}^{(1)}&=      \left[ \frac{1}{\left(
u-U\right) ^{2}}+\frac{2-u}{u\left( u-1\right) ^{2}}\right] \delta _{L}+%
\frac{U\left( U-1\right) }{u\left( u-1\right) \left( u-U\right) }d_{4}+V\left[ \frac{2}{\left( U-u\right) ^{3}}\epsilon_{L}+\frac{1}{u\left(
u-1\right) }c_{4}\right]     \, , \\
{6\over{c_M}}{\cal M}^{(0)}&={1\over{u^2}}{{1-\beta^2}\over 4}\, ,\\
 {6\over{c_M}}{\cal M}^{(1)}&=\left({{1}\over{(u-U)^2}}+{{1}\over{u(1-u)^2}} +{{1}\over{u(1-u)}}  \right) \epsilon_L +\left( {1\over{U-u}}+{1\over u}-{U\over{u(1-u)}}   \right)c_4 \, .
\end{split}
\ee
Evaluating the Wronskians in \ref{eq:Wapp} and \ref{eq:ziapp} and plugging the answer in \ref{eq:fprimeL} gives rise to solutions for $f^{\prime}_{i,j}(u)$. In order to obtain the solutions $X_i^{(1)}$ we would have to integrate the resulting expressions to obtain formulas for $f_{i,j}(u)$. However, we are only concerned with how the solutions of the differential equation rotate among each other when we move the current insertion around the light pair of primaries $L(x_3)L(x_4)$. This means we only need to compute residues of the integrand defining $f^{\prime}_{i,j}(u)$ at the points $u=1$ and $u=U$. Once this is done, we combine the results into the gauge invariant object $I_3$ defined in formula \ref{eq:I3def}. This finally yields
\be
\begin{split}
\sqrt{I_3}=& 4\pi \left[ V \left(  {{  U^{\beta}(\beta+2)+\beta-2     }\over{2\beta U(U^{\beta}-1)}} \epsilon_{\alpha}^2  +{{\beta-1 -U^{\beta}(\beta+1)(U^{\beta}+2\beta-2)}\over{\beta^2 (U^{\beta}-1)U^{\beta/2+1}}} \epsilon_{\alpha}\epsilon_L      \right)                                                 \right. \\
&\left. + {{U^{\beta}-1}\over{U^{\beta/2-1} \beta^2}} \epsilon_{\alpha}d_4       +      {{U^{\beta}(\beta+1)+\beta-1}\over{\beta^2 U^{\beta/2}}} \delta_L \epsilon_{\alpha}  \right.\\
&\left.     +\delta_H   \left(   {{(2\beta+\beta(U^{\beta}-1))\log U-2(U^{\beta}-1)}\over{(U^{\beta}-1)\beta^3}}\epsilon_{\alpha}^2   +2{{U^{2\beta}-1-2\beta U^{\beta}\log U}\over{\beta^3(U^{\beta}-1)U^{\beta/2}}}\epsilon_{\alpha}\epsilon_L            \right)             \right]^{{1\over 2}}\, .
\end{split}
\ee
Comparing this result with the one we found when studying the three-point function in equation \ref{eq:I3app} yields a solution for $d_4$, which is the one we presented in formula \ref{eq:d4}.


\section{Limiting analysis}\label{app:limit}
The BMS$_3$ algebra can be obtained as a result of a non-relativistic contraction of the Virasoro algebras of a conformal field theory in two dimensions. The contraction in space-time reads
\be
t\rightarrow t\, ,\quad \text{and}\quad x\rightarrow \epsilon x\, .
\ee
At the level of the algebra we have
\be
L_n = {\cal L}_n+\bar{{\cal L}}_n\, ,\quad \text{and}\quad M_n=\epsilon \left(   {\cal L}_n-\bar{{\cal L}}_n  \right)\, .
\ee
This means that eigenvalues of the BMS operators are related to eigenvalues of BMS operators as 
\be
\Delta = h+\bar{h}\, , \quad \text{and}\quad \xi=\epsilon \left(  h-\bar{h}  \right)\, .
\ee
In a conformal field theory, a finite conformal map manifests itself in the coordinates as
\be
z\rightarrow f(z)\, , \quad \text{and}\quad \bar{z}\rightarrow \bar{f}(\bar{z})\, ,
\ee
where $z=t+x$ and $\bar{z}=t-x$. Taking the non-relativistic limit we obtain
\be
t\rightarrow P(t)\, , \quad \text{and}\quad x\rightarrow G(t) +x P'(t) \, .
\ee
where $P$ and $G$ are related to the conformal maps as  follows
\be
P(t)={{f(t)+\bar{f}(t)}\over 2}\, , \quad \text{and}\quad G(t)={{f(t)-\bar{f}(t)}\over {2\epsilon}}\, .
\ee
In this appendix we will explore the transformation laws of primaries and currents, as well as the explicit results for BMS blocks by taking the limit explained above.

\subsection{Primaries and Currents}\label{sec:limit}
Primaries in a conformal field theory transform in the following way
\be
\phi'(z,\bar{z})=f'(z)^h \bar{f}'(\bar{z})^{\bar{h}} \phi(f(z),\bar{f}(\bar{z}))\, .
\ee
Replacing the coordinates, the conformal maps and the eigenvalues by their relations to $P$, $G$, and $\epsilon$ and expansing in $\epsilon$ we recover
\be\label{eq:primTransfCFT}
\phi'(u,v)=\left(P'\right)^{\Delta} e^{\xi \left(  {{G'}\over{P'}}+v{{P''}\over{P'}}   \right)}\phi (P,vP'+G)\, ,
\ee
where we have replaced $t\leftrightarrow u$ and $x\leftrightarrow v$. The same techniques can be used to compute the transformation laws of the stress tensors. We start with 
\be
T'(z)=f'(z) ^2 T(f(z))+{{c}\over{12}}S(f(z),z)\, ,
\ee
and similarly for the barred component. The central charges $c$ and $\bar{c}$ are related to $c_L$ and $c_M$ as
\be
c={{c_L+\epsilon^{-1}c_M}\over{2}}\, , \quad \text{and}\quad  \bar{c}={{c_L-\epsilon^{-1}c_M}\over{2}}\, .
\ee
We now replace coordinates and maps with their non-relativistic limits and obtain the same expressions we found in section \ref{sec:currents}.

\subsection{The BMS$_3$ block from the Virasoro block }
The BMS$_3$ block is by definition only dependent on the structure of the BMS algebra. BMS$_3$ blocks are nothing but a basis of BMS invariant functions one can use to expand a four-point correlation function. The BMS$_3$ algebra can be constructed as a non-relativistic contraction of the Virasoro algebra \cite{Bagchi:2017cpu}. We thus expect that a particular limit of a Virasoro block will result in the BMS blocks we have computed throughout this note. The non-relativistic contraction has already been used above  in order to check the transformation laws of BMS primaries and currents. In this appendix we will apply such contraction to the Virasoro block of four primary operators in a conformal field theory. We start with the expression for the block involving two heavy and two light CFT primaries. The CFT correlator we consider is
\be
\langle {\cal O}_H(\infty,\infty) {\cal O}_H (0,0) {\cal O}_L(z,\bar{z}) {\cal O}_L (1,1) \rangle\, .
\ee
The Virasoro blocks in which this correlator can be expanded factorize holomorphically
\be
{\cal V} (z,\bar{z}) = {\cal F}(h_i,h_p,c,z-1) \times \bar{{\cal F}}(\bar{h}_i,\bar{h}_p,\bar{c},\bar{z}-1) \, ,
\ee
with
\be\label{eq:CFTBlock}
{\cal F}(h_i,h_p,c;z-1) =z^{(\beta_{\text{CFT}}-1)h_L}  (1-z^{\beta_{\text{CFT}}})^{h_{\alpha}-2h_L} {}_2 F_1 \left(   h_{\alpha} ,h_{\alpha} ,2h_{\alpha};1-z^{\beta_{\text{CFT}}}  \right)\, ,
\ee
and
\be
\beta_{\text{CFT}}=\sqrt{1-{{24h_H}\over{c}}}\, .
\ee
We will only compute the contraction of the holomorphic part of the block, and deduce the anti-holomorphic part by making the appropriate replacements. We take the limit by making the following replacements
\be\label{eq:replacements}
\begin{split}
h_{L,H,\alpha}&={{\Delta_{L,H,\alpha}}\over{2}}+{{\xi_{L,H,\alpha}}\over{2\epsilon}}\, , \\
c&={{c_L+c_M/\epsilon}\over{2}}\, \\
z&=u+\epsilon v \, .
\end{split}
\ee
Note that under these replacements, we have
\be
\beta_{\text{CFT}}=\sqrt{1-{{24\xi_H}\over{c}}}-2 \epsilon{{6\Delta_H}\over{c_M}}{1\over{\sqrt{1-{{24\xi_H}\over{c}}}}}+{\cal O}(\epsilon )^2 = \beta-2{{\delta_H}\over{\beta}}\epsilon+{\cal O}(\epsilon)^2\, .
\ee
With these replacements at hand, we can write the prefactors multiplying the hypergeometric function in equation \ref{eq:CFTBlock} as
\be\label{eq:prefactor}
\begin{split}
z^{(\beta_{\text{CFT}}-1)h_L}  (1-z^{\beta_{\text{CFT}}})^{h_{\alpha}-2h_L} \rightarrow& u^{{1\over 2}(\beta-1)\Delta_L} (1-u^{\beta})^{{1\over 2}(\Delta_{\alpha}-2\Delta_L)} e^{{v\over 2}\left(  {{\beta u^{\beta-1}}\over{u^{\beta}-1}}   \xi_{\alpha}      -{{\beta-1+u^{\beta}(\beta+1)}\over{u(u^{\beta}-1)}}\xi_L      \right)}\\ 
&\times e^{{{\delta_H}\over{\beta}} \log u \left(   {{u^{\beta}+1}\over{u^{\beta}-1}}\xi_L     {{u^{\beta}}\over{1-u^{\beta}}}  \xi_{\alpha}   \right)} e^{{1\over{\epsilon}}   \left(   \xi_{\alpha}\log\sqrt{1-u^{\beta}}+\xi_L \log{{u^{(\beta-1)/2}}\over{1-u^{\beta}}}      \right)                        }\, .
\end{split}
\ee
We also need to take the limit of the ${}_2 F_1$ hypergeometric function. We will make use of the integral formula
\be\label{eq:2F1}
{}_2 F_1(a,b,c;\chi) ={{\int_{0}^{1} p^{b-1}(1-p)^{c-b-1}(1-p\chi)^{-a}dp}\over{B(b,c-b)}}\, ,
\ee
where $B(b,c-b)$ is the incomplete Beta function, which also has an integral formula
\be
B(b,c-b)=\int_0^1 p^{b-1}(1-p)^{c-b-1} dp\, .
\ee
Using the replacements \ref{eq:replacements} leads to the following integral for the numerator in \ref{eq:2F1}
\be
\int_{0}^{1} p^{b-1}(1-p)^{c-b-1}(1-p\chi)^{-a}dp=\int_{0}^{1} dp   \, e^{{1\over{\epsilon}} S(p)} g(p) \, ,
\ee
with
\be
\begin{split}
S(p)&=\xi_{\alpha} \log\sqrt{    {{p(1-p)}\over{ 1-p(1-u^{\beta})}}       } \, , \\
g(p)&=e^{-{v\over 2} \beta \xi_{\alpha}  {{p u^{\beta-1}}\over{ 1-p(1-u^{\beta})    }}     +{{\delta_H}\over{\beta}} \xi_{\alpha} \log u {{p u^{\beta}}\over{1-p(1-u^{\beta})}}   }\left[ p(1-p)\right]^{{{\Delta}\over 2}-1} \left[  1-p(1-u^{\beta})  \right]^{-{{\Delta}\over 2}}
\end{split}
\ee
As we are taking the limit $\epsilon\rightarrow 0$, the integral can be approximated by a saddle point. We thus solve $S'(p)=0$ and find that the only solution in the interval $p\in [0,1]$ is
\be
p_{\star}={1\over{1+u^{{\beta}\over 2}}}\, .
\ee
The integral is then
\be\label{eq:Numerator}
\begin{split}
\int_{0}^{1} p^{b-1}(1-p)^{c-b-1}(1-p\chi)^{-a}dp\approx &\sqrt{{{2\pi \epsilon}\over{-S''(p_{\star})}}} e^{{1\over {\epsilon}} S(p_{\star})} g(p_{\star})   \, \\
=&\sqrt{{{2\pi \epsilon}\over{\xi_{\alpha}}}} e^{-{1\over {\epsilon}} \xi_{\alpha} \log \left(  1+u^{{{\beta}\over 2}}  \right)} u^{-{{\beta}\over 4}} \left(  1+u^{{{\beta}\over 2}}  \right)^{1-\Delta_{\alpha}} \\
&\times e^{  -{v\over 2} \beta \xi_{\alpha} {1\over {u   \left(  1+u^{-{{\beta}\over 2}}  \right) }}      +{{\delta_H}\over {\beta}}  \xi_{\alpha} \log u {1\over{\left(  1+u^{-{{\beta}\over 2}}  \right)}}           }\, .
\end{split}
\ee
The integral in the denominator of expression \ref{eq:2F1} can be computed in the same fashion. In this case the saddle point is at $p_{\star}=1/2$, and leads to the following expression
\be\label{eq:Denominator}
\int_0^1 p^{b-1}(1-p)^{c-b-1} dp\approx  \sqrt{{2\pi \epsilon}\over{4\xi_{\alpha}}} e^{-{1\over{\epsilon}}\xi_{\alpha}\log 2} {1\over 2^{\Delta_{\alpha}-2}}\, .
\ee
The resulting expression for the holomorphic block can be obtained by putting together expressions \ref{eq:prefactor}, \ref{eq:Numerator}, and \ref{eq:Denominator}. The expression for the anti-holomorphic block can be obtained then by replacing $\xi_{L,\alpha}\rightarrow -\xi_{L,\alpha}$, $v\rightarrow -v$, $\delta_H\rightarrow -\delta_H$. Once this is done, all exponentials that diverge in the $\epsilon \rightarrow 0$ limit cancel out and we are left with the final expression
\be\label{eq:limitresult}\begin{split}
{\cal V} (z,\bar{z}) \rightarrow& \left( {{1-u^{{\beta}\over{2}}}\over{1+u^{{\beta}\over{2}}}}  \right)^{\Delta_{\alpha}}   \left( {{   u^{\beta-1}    }\over{  (1-u^{\beta})^2    }}   \right)^{\Delta_L} e^{v\left(  {{\beta u^{{{\beta}\over 2}}}\over{u(u^{\beta}-1)}}\xi_{\alpha} - {{u^{\beta}(\beta+1)+\beta-1}\over{u(u^{\beta}-1)}}\xi_L \right)
+\delta_H \log u \left(     {{2u^{{\beta}\over 2}}\over{\beta (u^{\beta}-1)}}    \xi_{\alpha}            +  {{2(u^{\beta}+1)}\over{\beta(1-u^{\beta})}} \xi_L                    \right)}\\
&\times 2^{2\Delta_{\alpha}} u^{-{{\beta}\over 2}} \left( 1+u^{{{\beta}\over 2}}     \right)^2\, .
\end{split}
\ee
The result agrees with the formulas found through the application of the monodromy method in a BMSFT, as well as the holographic computations in a flat space cosmological solution. The only difference between this result and the results found in previous sections are the factors in the second line of \ref{eq:limitresult}. This is expected, as these factors are subleading in the small $\epsilon_{L,\alpha}$, $\delta_{L,\alpha}$ limit, and so they do not appear in our monodromy calculations or their holographic counterparts. It is also worth pointing out that the computations done in previous sections have been done in the context of a possible flat-space/BMSFT duality without any reference to AdS/CFT.

--

\bibliographystyle{JHEP}
\bibliography{refs}

\providecommand{\href}[2]{#2}\begingroup\raggedright\begin{thebibliography}{10}

\bibitem{Hijano:2017eii}
E.~Hijano and C.~Rabideau, \emph{{Holographic Entanglement and Poincare blocks
  in three dimensional flat space}},
  \href{https://arxiv.org/abs/1712.07131}{{\tt 1712.07131}}.

\bibitem{tHooft:1999rgb}
G.~'t~Hooft, \emph{{The Holographic principle: Opening lecture}},
  \href{http://dx.doi.org/10.1142/9789812811585_0005}{\emph{Subnucl. Ser.} {\bf
  37} (2001) 72--100}, [\href{https://arxiv.org/abs/hep-th/0003004}{{\tt
  hep-th/0003004}}].

\bibitem{Susskind:1994vu}
L.~Susskind, \emph{{The World as a hologram}},
  \href{http://dx.doi.org/10.1063/1.531249}{\emph{J. Math. Phys.} {\bf 36}
  (1995) 6377--6396}, [\href{https://arxiv.org/abs/hep-th/9409089}{{\tt
  hep-th/9409089}}].

\bibitem{Bagchi:2010eg}
A.~Bagchi, \emph{{Correspondence between Asymptotically Flat Spacetimes and
  Nonrelativistic Conformal Field Theories}},
  \href{http://dx.doi.org/10.1103/PhysRevLett.105.171601}{\emph{Phys. Rev.
  Lett.} {\bf 105} (2010) 171601}, [\href{https://arxiv.org/abs/1006.3354}{{\tt
  1006.3354}}].

\bibitem{Bagchi:2012cy}
A.~Bagchi and R.~Fareghbal, \emph{{BMS/GCA Redux: Towards Flatspace Holography
  from Non-Relativistic Symmetries}},
  \href{http://dx.doi.org/10.1007/JHEP10(2012)092}{\emph{JHEP} {\bf 10} (2012)
  092}, [\href{https://arxiv.org/abs/1203.5795}{{\tt 1203.5795}}].

\bibitem{Barnich:2010eb}
G.~Barnich and C.~Troessaert, \emph{{Aspects of the BMS/CFT correspondence}},
  \href{http://dx.doi.org/10.1007/JHEP05(2010)062}{\emph{JHEP} {\bf 05} (2010)
  062}, [\href{https://arxiv.org/abs/1001.1541}{{\tt 1001.1541}}].

\bibitem{Bondi:1962px}
H.~Bondi, M.~G.~J. van~der Burg and A.~W.~K. Metzner, \emph{{Gravitational
  waves in general relativity. 7. Waves from axisymmetric isolated systems}},
  \href{http://dx.doi.org/10.1098/rspa.1962.0161}{\emph{Proc. Roy. Soc. Lond.}
  {\bf A269} (1962) 21--52}.

\bibitem{Sachs:1962zza}
R.~Sachs, \emph{{Asymptotic symmetries in gravitational theory}},
  \href{http://dx.doi.org/10.1103/PhysRev.128.2851}{\emph{Phys. Rev.} {\bf 128}
  (1962) 2851--2864}.

\bibitem{Srednicki:1995pt}
M.~Srednicki, \emph{{Thermal fluctuations in quantized chaotic systems}},
  \href{http://dx.doi.org/10.1088/0305-4470/29/4/003}{\emph{J. Phys.} {\bf A29}
  (1996) L75--L79}, [\href{https://arxiv.org/abs/chao-dyn/9511001}{{\tt
  chao-dyn/9511001}}].

\bibitem{PhysRevA.43.2046}
J.~M. Deutsch, \emph{Quantum statistical mechanics in a closed system},
  \href{http://dx.doi.org/10.1103/PhysRevA.43.2046}{\emph{Phys. Rev. A} {\bf
  43} (Feb, 1991) 2046--2049}.

\bibitem{Lashkari:2016vgj}
N.~Lashkari, A.~Dymarsky and H.~Liu, \emph{{Eigenstate Thermalization
  Hypothesis in Conformal Field Theory}},
  \href{http://dx.doi.org/10.1088/1742-5468/aab020}{\emph{J. Stat. Mech.} {\bf
  1803} (2018) 033101}, [\href{https://arxiv.org/abs/1610.00302}{{\tt
  1610.00302}}].

\bibitem{Fitzpatrick:2016mjq}
A.~L. Fitzpatrick and J.~Kaplan, \emph{{On the Late-Time Behavior of Virasoro
  Blocks and a Classification of Semiclassical Saddles}},
  \href{http://dx.doi.org/10.1007/JHEP04(2017)072}{\emph{JHEP} {\bf 04} (2017)
  072}, [\href{https://arxiv.org/abs/1609.07153}{{\tt 1609.07153}}].

\bibitem{Fitzpatrick:2015zha}
A.~L. Fitzpatrick, J.~Kaplan and M.~T. Walters, \emph{{Virasoro Conformal
  Blocks and Thermality from Classical Background Fields}},
  \href{http://dx.doi.org/10.1007/JHEP11(2015)200}{\emph{JHEP} {\bf 11} (2015)
  200}, [\href{https://arxiv.org/abs/1501.05315}{{\tt 1501.05315}}].

\bibitem{Faulkner:2017hll}
T.~Faulkner and H.~Wang, \emph{{Probing beyond ETH at large $c$}},
  \href{https://arxiv.org/abs/1712.03464}{{\tt 1712.03464}}.

\bibitem{Bagchi:2017cpu}
A.~Bagchi, M.~Gary and Zodinmawia, \emph{{The nuts and bolts of the BMS
  Bootstrap}}, \href{http://dx.doi.org/10.1088/1361-6382/aa8003}{\emph{Class.
  Quant. Grav.} {\bf 34} (2017) 174002},
  [\href{https://arxiv.org/abs/1705.05890}{{\tt 1705.05890}}].

\bibitem{Bagchi:2016geg}
A.~Bagchi, M.~Gary and Zodinmawia, \emph{{Bondi-Metzner-Sachs bootstrap}},
  \href{http://dx.doi.org/10.1103/PhysRevD.96.025007}{\emph{Phys. Rev.} {\bf
  D96} (2017) 025007}, [\href{https://arxiv.org/abs/1612.01730}{{\tt
  1612.01730}}].

\bibitem{Gurarie:1993xq}
V.~Gurarie, \emph{{Logarithmic operators in conformal field theory}},
  \href{http://dx.doi.org/10.1016/0550-3213(93)90528-W}{\emph{Nucl. Phys.} {\bf
  B410} (1993) 535--549}, [\href{https://arxiv.org/abs/hep-th/9303160}{{\tt
  hep-th/9303160}}].

\bibitem{Harlow:2011ny}
D.~Harlow, J.~Maltz and E.~Witten, \emph{{Analytic Continuation of Liouville
  Theory}}, \href{http://dx.doi.org/10.1007/JHEP12(2011)071}{\emph{JHEP} {\bf
  12} (2011) 071}, [\href{https://arxiv.org/abs/1108.4417}{{\tt 1108.4417}}].

\bibitem{Anous:2016kss}
T.~Anous, T.~Hartman, A.~Rovai and J.~Sonner, \emph{{Black Hole Collapse in the
  1/c Expansion}}, \href{http://dx.doi.org/10.1007/JHEP07(2016)123}{\emph{JHEP}
  {\bf 07} (2016) 123}, [\href{https://arxiv.org/abs/1603.04856}{{\tt
  1603.04856}}].

\bibitem{deBoer:2014sna}
J.~de~Boer, A.~Castro, E.~Hijano, J.~I. Jottar and P.~Kraus, \emph{{Higher spin
  entanglement and $ {\mathcal{W}}_{\mathrm{N}} $ conformal blocks}},
  \href{http://dx.doi.org/10.1007/JHEP07(2015)168}{\emph{JHEP} {\bf 07} (2015)
  168}, [\href{https://arxiv.org/abs/1412.7520}{{\tt 1412.7520}}].

\bibitem{Barnich:2006av}
G.~Barnich and G.~Compere, \emph{{Classical central extension for asymptotic
  symmetries at null infinity in three spacetime dimensions}},
  \href{http://dx.doi.org/10.1088/0264-9381/24/5/F01,
  10.1088/0264-9381/24/11/C01}{\emph{Class. Quant. Grav.} {\bf 24} (2007)
  F15--F23}, [\href{https://arxiv.org/abs/gr-qc/0610130}{{\tt gr-qc/0610130}}].

\bibitem{Barnich:2015uva}
G.~Barnich and B.~Oblak, \emph{{Notes on the BMS group in three dimensions: II.
  Coadjoint representation}},
  \href{http://dx.doi.org/10.1007/JHEP03(2015)033}{\emph{JHEP} {\bf 03} (2015)
  033}, [\href{https://arxiv.org/abs/1502.00010}{{\tt 1502.00010}}].

\bibitem{Oblak:2016eij}
B.~Oblak, \emph{{BMS Particles in Three Dimensions}}.
\newblock PhD thesis, Brussels U., 2016.
\newblock \href{https://arxiv.org/abs/1610.08526}{{\tt 1610.08526}}.
\newblock 10.1007/978-3-319-61878-4.

\bibitem{Bagchi:2009ca}
A.~Bagchi and I.~Mandal, \emph{{On Representations and Correlation Functions of
  Galilean Conformal Algebras}},
  \href{http://dx.doi.org/10.1016/j.physletb.2009.04.030}{\emph{Phys. Lett.}
  {\bf B675} (2009) 393--397}, [\href{https://arxiv.org/abs/0903.4524}{{\tt
  0903.4524}}].

\bibitem{Bagchi:2009pe}
A.~Bagchi, R.~Gopakumar, I.~Mandal and A.~Miwa, \emph{{GCA in 2d}},
  \href{http://dx.doi.org/10.1007/JHEP08(2010)004}{\emph{JHEP} {\bf 08} (2010)
  004}, [\href{https://arxiv.org/abs/0912.1090}{{\tt 0912.1090}}].

\bibitem{Belavin:1984vu}
A.~A. Belavin, A.~M. Polyakov and A.~B. Zamolodchikov, \emph{{Infinite
  Conformal Symmetry in Two-Dimensional Quantum Field Theory}},
  \href{http://dx.doi.org/10.1016/0550-3213(84)90052-X}{\emph{Nucl. Phys.} {\bf
  B241} (1984) 333--380}.

\bibitem{Fitzpatrick:2014vua}
A.~L. Fitzpatrick, J.~Kaplan and M.~T. Walters, \emph{{Universality of
  Long-Distance AdS Physics from the CFT Bootstrap}},
  \href{http://dx.doi.org/10.1007/JHEP08(2014)145}{\emph{JHEP} {\bf 08} (2014)
  145}, [\href{https://arxiv.org/abs/1403.6829}{{\tt 1403.6829}}].

\bibitem{Hartman:2013mia}
T.~Hartman, \emph{{Entanglement Entropy at Large Central Charge}},
  \href{https://arxiv.org/abs/1303.6955}{{\tt 1303.6955}}.

\bibitem{Hogervorst:2016itc}
M.~Hogervorst, M.~Paulos and A.~Vichi, \emph{{The ABC (in any D) of Logarithmic
  CFT}}, \href{http://dx.doi.org/10.1007/JHEP10(2017)201}{\emph{JHEP} {\bf 10}
  (2017) 201}, [\href{https://arxiv.org/abs/1605.03959}{{\tt 1605.03959}}].

\bibitem{Cardy:1999zp}
J.~L. Cardy, \emph{{Logarithmic correlations in quenched random magnets and
  polymers}},  \href{https://arxiv.org/abs/cond-mat/9911024}{{\tt
  cond-mat/9911024}}.

\bibitem{Caux:1995nm}
J.~S. Caux, I.~I. Kogan and A.~M. Tsvelik, \emph{{Logarithmic operators and
  hidden continuous symmetry in critical disordered models}},
  \href{http://dx.doi.org/10.1016/0550-3213(96)00118-6}{\emph{Nucl. Phys.} {\bf
  B466} (1996) 444--462}, [\href{https://arxiv.org/abs/hep-th/9511134}{{\tt
  hep-th/9511134}}].

\bibitem{Flohr:1997wm}
M.~A.~I. Flohr, \emph{{Singular vectors in logarithmic conformal field
  theories}},
  \href{http://dx.doi.org/10.1016/S0550-3213(97)00012-6}{\emph{Nucl. Phys.}
  {\bf B514} (1998) 523--552},
  [\href{https://arxiv.org/abs/hep-th/9707090}{{\tt hep-th/9707090}}].

\bibitem{Flohr:2000mc}
M.~Flohr, \emph{{Null vectors in logarithmic conformal field theory}},
  \href{https://arxiv.org/abs/hep-th/0009137}{{\tt hep-th/0009137}}.

\bibitem{Jiang:2017ecm}
H.~Jiang, W.~Song and Q.~Wen, \emph{{Entanglement Entropy in Flat Holography}},
  \href{http://dx.doi.org/10.1007/JHEP07(2017)142}{\emph{JHEP} {\bf 07} (2017)
  142}, [\href{https://arxiv.org/abs/1706.07552}{{\tt 1706.07552}}].

\bibitem{Hijano:2015rla}
E.~Hijano, P.~Kraus and R.~Snively, \emph{{Worldline approach to semi-classical
  conformal blocks}},
  \href{http://dx.doi.org/10.1007/JHEP07(2015)131}{\emph{JHEP} {\bf 07} (2015)
  131}, [\href{https://arxiv.org/abs/1501.02260}{{\tt 1501.02260}}].

\bibitem{Hijano:2015zsa}
E.~Hijano, P.~Kraus, E.~Perlmutter and R.~Snively, \emph{{Witten Diagrams
  Revisited: The AdS Geometry of Conformal Blocks}},
  \href{http://dx.doi.org/10.1007/JHEP01(2016)146}{\emph{JHEP} {\bf 01} (2016)
  146}, [\href{https://arxiv.org/abs/1508.00501}{{\tt 1508.00501}}].

\bibitem{Cornalba:2002eg}
L.~Cornalba, M.~S. Costa and C.~Kounnas, \emph{{A new cosmological scenario in
  string theory}},  in \emph{{Supersymmetry and unification of fundamental
  interactions. Proceedings, 10th International Conference, SUSY'02, Hamburg,
  Germany, June 17-23, 2002}}, pp.~1315--1324, 2002.

\bibitem{Cornalba:2003kd}
L.~Cornalba and M.~S. Costa, \emph{{Time dependent orbifolds and string
  cosmology}}, \href{http://dx.doi.org/10.1002/prop.200310123}{\emph{Fortsch.
  Phys.} {\bf 52} (2004) 145--199},
  [\href{https://arxiv.org/abs/hep-th/0310099}{{\tt hep-th/0310099}}].

\bibitem{Barnich:2012aw}
G.~Barnich, A.~Gomberoff and H.~A. Gonzalez, \emph{{The Flat limit of three
  dimensional asymptotically anti-de Sitter spacetimes}},
  \href{http://dx.doi.org/10.1103/PhysRevD.86.024020}{\emph{Phys. Rev.} {\bf
  D86} (2012) 024020}, [\href{https://arxiv.org/abs/1204.3288}{{\tt
  1204.3288}}].

\bibitem{Anous:2017tza}
T.~Anous, T.~Hartman, A.~Rovai and J.~Sonner, \emph{{From Conformal Blocks to
  Path Integrals in the Vaidya Geometry}},
  \href{http://dx.doi.org/10.1007/JHEP09(2017)009}{\emph{JHEP} {\bf 09} (2017)
  009}, [\href{https://arxiv.org/abs/1706.02668}{{\tt 1706.02668}}].

\bibitem{Castro:2014tta}
A.~Castro, S.~Detournay, N.~Iqbal and E.~Perlmutter, \emph{{Holographic
  entanglement entropy and gravitational anomalies}},
  \href{http://dx.doi.org/10.1007/JHEP07(2014)114}{\emph{JHEP} {\bf 07} (2014)
  114}, [\href{https://arxiv.org/abs/1405.2792}{{\tt 1405.2792}}].

\end{thebibliography}\endgroup
\end{document}